\documentclass[aps, prl, superscriptaddress, reprint, twocolumn, amsmath, amssymb, a4, notitlepage]{revtex4-1}
\usepackage{graphicx}
\usepackage{amsmath,amssymb}
\usepackage{color}
\usepackage{mathrsfs}
\usepackage{physics}
\usepackage[utf8]{inputenc}

\usepackage{mathptmx}
\usepackage{newtxmath}
\usepackage{newtxtext}

\begin{document}
\title{
Collective quantum memory activated by a driven central spin
}
\author{Emil V. Denning}
\affiliation{Department of Photonics Engineering, Technical University of Denmark, 2800 Kgs. Lyngby, Denmark}
\affiliation{Cavendish Laboratory, University of Cambridge, JJ Thomson Avenue, Cambridge CB3 0HE, United Kingdom}

\author{Dorian A. Gangloff}
\affiliation{Cavendish Laboratory, University of Cambridge, JJ Thomson Avenue, Cambridge CB3 0HE, United Kingdom}

\author{Mete Atat\"ure}
\affiliation{Cavendish Laboratory, University of Cambridge, JJ Thomson Avenue, Cambridge CB3 0HE, United Kingdom}

\author{Jesper M\o rk}
\affiliation{Department of Photonics Engineering, Technical University of Denmark, 2800 Kgs. Lyngby, Denmark}

\author{Claire Le Gall}
\email[Electronic address: ]{cl538@cam.ac.uk}
\affiliation{Cavendish Laboratory, University of Cambridge, JJ Thomson Avenue, Cambridge CB3 0HE, United Kingdom}

\date{\today}

\begin{abstract}
Coupling a qubit coherently to an ensemble is the basis for collective quantum memories. A driven quantum dot can deterministically excite low-energy collective modes of a nuclear spin ensemble in the presence of lattice strain. We propose to gate a quantum state transfer between this central electron and these low-energy excitations -- spin waves -- in the presence of a strong magnetic field, where the nuclear coherence time is long. We develop a microscopic theory capable of calculating the exact time evolution of the strained electron-nuclear system. With this, we evaluate the operation of quantum state storage and show that fidelities up to 90\% can be reached with a modest nuclear polarisation of only 50\%. These findings demonstrate that strain-enabled nuclear spin waves are a highly suitable candidate for quantum memory.

\end{abstract}
	\maketitle

\paragraph{Introduction \textemdash}
\label{sec:introduction}
Quantum memory working in conjunction with a computational qubit is a central element in fault-tolerant quantum computing and communication strategies~\cite{briegel1998quantum,kimble2008quantum}. 
To name a few prominent examples, quantum memories have been demonstrated using collective states of atomic ensembles to store a photonic qubit~\cite{kozhekin2000quantum,julsgaard2004experimental,choi2010entanglement,hedges2010efficient,zhao2009long}, cold ions, where the decoherence-free subspace of a local ion pair acts as memory for a single ion qubit~\cite{kielpinski2001decoherence,langer2005long} and nitrogen vacancy centres in diamond, where the electronic spin state can be written into a single proximal nuclear spin~\cite{dutt2007quantum,fuchs2011quantum,taminiau2014universal}. For semiconductor quantum dots, the mesoscopic spin environment comprising $\sim 10^4-10^5$ nuclei is a candidate for a collective quantum memory that can store the electronic spin state~\cite{taylor2003long,taylor2003controlling,kurucz2009qubit}, with the promise of coherence times reaching milliseconds~\cite{chekhovich2015suppression}. A strategy for electron--nuclear state transfer is based on flip-flops generated by the collinear hyperfine interaction~\cite{taylor2003long}. A consequence of this interaction scheme brings about opposing requirements: Vanishing electron spin splitting during the state transfer versus large electron spin splitting to polarise and stabilise the nuclear coherence~\cite{wust2016role}. An alternative approach is to use collective nuclear spin wave excitations that have recently been observed under a strong magnetic field in the form of a coherently distributed single nuclear spin excitation~\cite{gangloff2019magnon} through an effective non-collinear hyperfine interaction~\cite{hogele2012dynamic}. In this paper, we propose a protocol for quantum memory based on this interaction, which in equilibrium is suppressed by a strong static magnetic field, but can be controllably switched on for a finite time by driving the qubit out of equilibrium.

\begin{figure}
  \centering
  \includegraphics[width=\columnwidth]{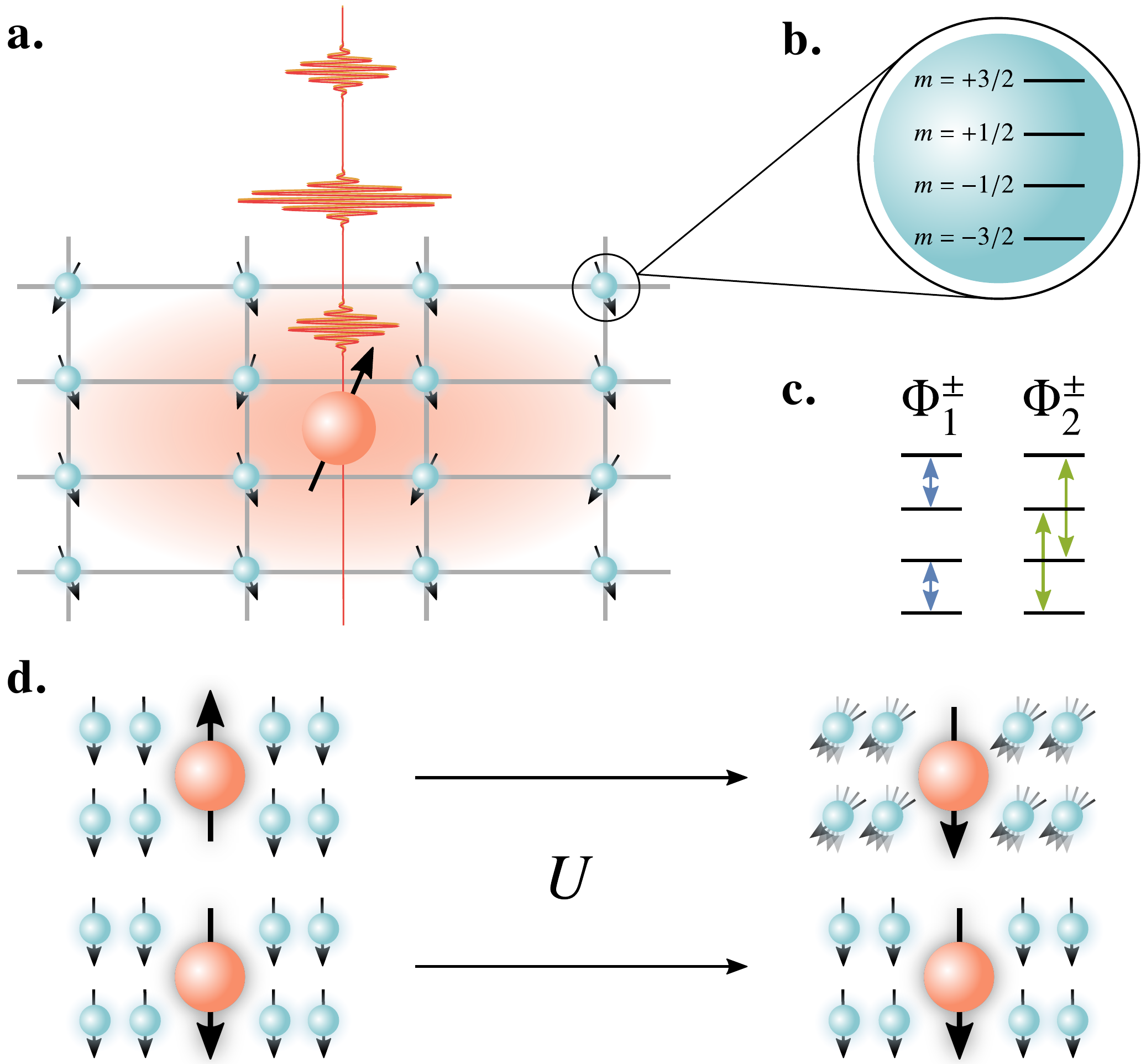}
  \caption{{\bf a.} System schematic comprising a pulse-driven electron coupled to a mesoscopic bath of nuclear spins
{\bf b.} Structure of the high-spin ($I>1/2$) nuclei, here shown for $I=3/2$. {\bf c.} Spin transitions between states in b. generated by the non-collinear processes $\Phi_1^\pm$ and $\Phi_2^\pm$
{\bf d.} In the memory write-in process, the stimulated electron--nuclear interaction flips the electron and generates a nuclear spin wave conditionally on the electron spin, thereby transferring the electron state to the nuclear ensemble. Here, the time-evolution operator, $U=e^{-i\mathcal{H}_I t}$, corresponds to the evolution in Eq.~\eqref{eq:ideal-dyn} at time $t=\pi/(2g_\zeta)$.
}
  \label{fig:1-1}
\end{figure}

The non-collinear hyperfine interaction responsible for qubit-controlled spin wave excitation originates from strain. In a strained lattice (cf. Figure~\ref{fig:1-1}a), the induced electric field gradient couples to the quadrupole moment of the nuclei, thereby tilting the nuclear spin quantisation axis away from that dictated by the magnetic field (defining the $z$-axis). 
This strain-induced mixing of the Zeeman eigenstates allows otherwise forbidden nuclear transitions that can be accessed by the electron through the hyperfine interaction, $H_\mathrm{hf}=\sum_j 2A^jI_z^j S_z$ ($S_z$ and $I_z^j$ are electronic and $j$th nuclear spin-$z$ operators). The transitions are activated when the electron spin is driven -- magnetically or optically -- to bridge the excitation energy gap corresponding to the nuclear Zeeman energy, $\omega_\mathrm{Z}^\mathrm{n}$.

Equipped with an interaction mechanism that can be switched on and off, information can be controllably transferred from the electron to the nuclei by letting the two subsystems interact for a finite duration. 
This can be realised in multiple ways; for the current proposal, we consider a Hamiltonian engineering approach based electron spin rotations on the Bloch sphere with a sequence of fast pulses~\cite{schwartz2018robust}, which selectively enhances the collective electron--nuclear transitions and simultaneously cancels out slow noise from the nuclear Overhauser field. The electron spin rotations in the pulse sequence can be carried out using an all-optical Raman drive to obtain phase-controlled manipulation at Rabi frequencies far exceeding the nuclear Zeeman splitting and hyperfine fluctuations of the electron Zeeman energy~\cite{bodey2019ESR}.
The nuclear coherence time can be extended up to milliseconds by removing the electron from the quantum dot~\cite{wust2016role} or alternatively by decoupling of the Knight field through a simple electron spin echo sequence~\cite{yao2006theory}. Read-out of the nuclear memory is effectuated by once again driving the electron to turn on the interaction.

\paragraph{Electron--nuclear exchange mechanism \textemdash}
The Hamiltonian describing the quadrupolar coupling of the $N$ nuclear spins ($I>1/2$) is
\begin{align*}
H_\mathrm{Q}=\hspace{-0.1cm}\sum_{j=1}^N B_\mathrm{Q}[(I_x^j)^2\sin^2\theta+\frac{1}{2}(I_x^jI_z^j+I_z^jI_x^j)\sin2\theta + (I_z^j)^2\cos^2\theta],
\end{align*}
where $I_\alpha^j,\; \alpha=x,y,z$ are the spin operators of the $j$'th nucleus, $\theta$ is the tilt angle of the quadrupolar axis away from $z$ and $B_\mathrm{Q}$ is the quadrupolar interaction strength. The low-energy excitations of the system are obtained through a Schrieffer-Wolff transformation perturbative in $B_\mathrm{Q}/\omega_\mathrm{Z}^\mathrm{n}$, which replaces $H_\mathrm{Q}$ by $H_\mathrm{Q}^0+V_\mathrm{Q}'$. $H_\mathrm{Q}^0$ commutes with $I_z^j$ and $V_\mathrm{Q}'=S_z[\mathcal{A}_1(\Phi_1^++\Phi_1^-)+ \mathcal{A}_2(\Phi_2^+ +\Phi_2^-)]$ is a non-collinear hyperfine interaction~\cite{gangloff2019magnon,suppmat}. Here, $\Phi_\zeta^+ \; (\zeta=1,2)$ denotes the nuclear spin wave operators
\begin{align*}
\Phi_1^+&=\sum_ja_{j}[I_+^jI_z^j+I_z^jI_+^j], \;\;
  \Phi_2^+=\sum_j a_{j} (I_+^j)^2,
\end{align*}
with $\Phi_\zeta^-=(\Phi_\zeta^+)^\dagger$; thus $\Phi_\zeta\pm$ changes the net nuclear spin by $\pm\zeta$ as shown in Figure~\ref{fig:1-1}b-c. The overall strength of the interaction is given by $\mathcal{A}_1=\frac{1}{2}\sum_jA^jB_\mathrm{Q}\sin2\theta/\omega_\mathrm{Z}^\mathrm{n}$ and $\mathcal{A}_2=\frac{1}{2}\sum_jA^jB_\mathrm{Q}\sin^2\theta/\omega_\mathrm{Z}^\mathrm{n}$, and $a_j=A^j/\sum_{j'}A^{j'}$ are the normalised hyperfine coefficients. Due to the collective nature of the nuclear spin waves, the typical coupling rates seen in the dynamics of the system are not the bare $\mathcal{A}_\zeta$, but rather the collectively enhanced rates, $\sim \mathcal{A}_\zeta\sqrt{N}$. A quadrupolar coupling strength of $B_\mathrm{Q}=1.5\mathrm{\;MHz}$, which is typical of naturally occuring strain~\cite{bulutay2012quadrupolar} or which can be engineering in situ~\cite{yuan2018uniaxial,flisinski2010optically} will be considered here.

A nuclear spin transition corresponding to the action of $\Phi_\zeta^\pm$ costs an energy of $\zeta\omega_\mathrm{Z}^\mathrm{n}$, which in a strong magnetic field is considerably larger than $\mathcal{A}_\zeta\sqrt{N}$. Consequently, these processes are far off-resonance in equilibrium.
To switch the interaction on, we follow Ref.~\cite{schwartz2018robust} and consider the action of a pulse sequence on the electron spin, driving it with a set of short $S_x$ and $S_y$ pulses separated by a time interval, $\tau/4$. By setting $\tau=\ell\pi/(\zeta\omega_\mathrm{Z}^\mathrm{n})$, where $\ell$ is an odd integer, the coupling between the electron and the $\Phi_\zeta$-mode is resonantly enhanced, and the system will evolve under an effective flip-flop Hamiltonian~\cite{suppmat}
\begin{align}
  \label{eq:6}
  \mathcal{H}_I=\mathcal{A}_\zeta'(\Phi_\zeta^+S_-+\Phi_\zeta^-S_+),
\end{align}
where $S_\pm=S_x\pm iS_y$ are the electron spin-flip operators and $\mathcal{A}_\zeta'$ is a rescaled coupling rate taking its maximal value for $\ell=3$, where $\mathcal{A}_\zeta'=(2+\sqrt{2})/(3\pi)\mathcal{A}_\zeta$. When the nuclei are initialised in a fully polarised state, this Hamiltonian will create a nuclear spin wave and flip the electron spin conditionally on the electron spin state (see Figure~\ref{fig:1-1}d), thus forming the basis of information transfer between the electron and nuclear ensemble.

\begin{figure*}
  \centering
  \includegraphics[width=\textwidth]{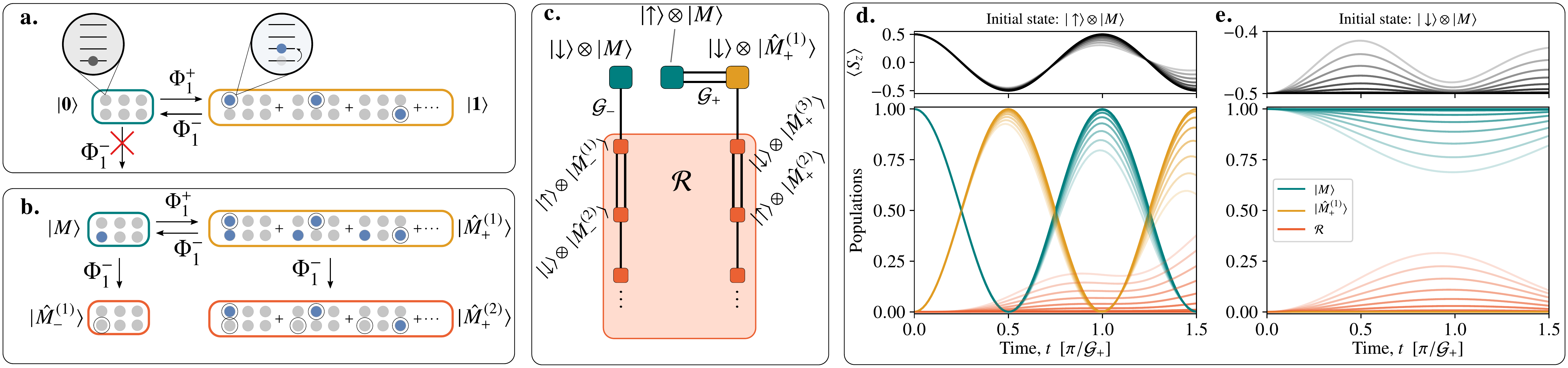}
  \caption{
{\bf a.} Collective spin wave excitation starting from a fully polarised state, where all nuclei are in the ground state (grey dots). The black circles emphasise the nuclei that have exchanged energy with the electron. The spin wave contains a single nuclear excitation (blue dots) distributed among all of the nuclei in a superposition. 
{\bf b.} Spin wave excitation from finitely polarised nuclear product state, $\ket{M}$, to target state, $\ket*{\hat{M}^{(1)}_+}$, where initially excited nuclei allow leakage transitions respectively to orthogonal states $\ket*{\hat{M}_-^{(1)}}$ and $\ket*{\hat{M}_+^{(2)}}$.
{\bf c.} Coupling structure in 1D mapping of nuclear state space. The initial state is a superposition of the two green states, and interactions couple these initial states to a 1D structure of states. Double lines signify the fast coupling rate $\mathcal{G}_+$, and single lines the slow rate, $\mathcal{G}_-$. For a fully polarised ensemble $\mathcal{G}_-=0$ and the three upper states remain isolated; at finite polarisation, this subspace is coupled to the residual chain of states, $\mathcal{R}$.
{\bf d.-e.} Dynamics for initialisation in electron states $\ket{\uparrow}$ and $\ket{\downarrow}$, respectively.
 Solid lines signify the fully polarised case, where $\mathcal{G}_-=0$. Lines with decreased opacity signify decreased polarisation and thus increased $\mathcal{G}_-$-rate, with the maximal value $\mathcal{G}_-/\mathcal{G}_+=0.3$. Line colors correspond to states in panel c. }
  \label{fig:1-2}
\end{figure*}

To see how this protocol turns nuclear spins into a quantum memory, we first consider a perfectly polarised nuclear bath (see Figure~\ref{fig:1-2}a). We write this nuclear state as $\ket{\mathbf{0}}=\ket{-I,\cdots,-I}$, and take the electron to be initialised in the state $\ket{\phi}=\alpha\ket{\uparrow}+\beta\ket{\downarrow}$, which we want to transfer to the nuclei. Because the nuclei are initialised in the ground state, no downwards transition are possible and $\Phi_\zeta^-\ket{\mathbf{0}}=0$. The excited nuclear state $\ket{\mathbf{1}}\propto\Phi_\zeta^+\ket{\mathbf{0}}$ is a distributed superposition of nuclear excitations (indicated by blue dots in Figure~\ref{fig:1-2}a), $\sum_j a_j\ket*{-I,\cdots,(-I+\zeta)_j,\cdots,-I}$. Crucially, when de-exciting the spin wave, the only downwards nuclear transitions available are those that were excited from the ground state, and thus $\Phi_\zeta^-\ket{\mathbf{1}}\propto\ket{\mathbf{0}}$. With these properties, it is straightforward to show that the system evolves within a three dimensional subspace as 
\begin{align}
\label{eq:ideal-dyn}
  \ket{\psi(t)}=\alpha\qty[\cos g_\zeta t \ket{\uparrow}\otimes\ket{\mathbf{0}}-i\sin g_{\zeta}t \ket{\downarrow}\otimes\ket{\mathbf{1}}] + \beta\ket{\downarrow}\otimes\ket{\mathbf{0}},
\end{align}
$g_\zeta=F_\zeta\mathcal{A}_\zeta'\sqrt{\sum_ja_j^2}$ (scaling as $\sqrt{N}$) is a collectively enhanced non-collinear coupling rate, where $F_1=(1-2I)\sqrt{2I},\; F_2=2\sqrt{I(2I-1)}$.
At time $t=\pi/(2g_\zeta)$ the electron--nuclear wavefunction separates, $\ket*{\psi(\pi/2g_\zeta)}=\ket{\downarrow}\otimes(-i\alpha\ket{\mathbf{0}}+\beta\ket{\mathbf{1}})$, and the electron state is identically transferred to the collective state of the nuclei to be stored.

\paragraph{Operation at partial nuclear polarisation \textemdash} A realistic implementation will initialise the nuclei in a partially polarised state~\cite{chekhovich2017measurement}, $\ket{M}=\ket{m_1,\cdots,m_N}$, such that there is a small number of lower energy states to scatter into, and thus $\Phi_\zeta^-\ket{M}\neq 0$ (see Figure~\ref{fig:1-2}b). Accordingly, the downwards transition $\ket{M}\rightarrow\Phi_\zeta^-\ket{M}$ is no longer forbidden as in the perfectly polarised case, but will take place with a rate, $\mathcal{G}_-$, which is slower than the upwards coupling rate, $\mathcal{G}_+$. Similarly, the downwards transition $\Phi_\zeta^-$ from the excited state $\Phi_\zeta^+\ket{M}$ does not lead back to $\ket{M}$ but mixes with other states generated by de-excitation of the initially unpolarised nuclei. This leads to dephasing of the spin wave mode serving as quantum memory. Nonetheless, the asymmetry of the coupling rates ($\mathcal{G}_+>\mathcal{G}_-$) makes it possible to operate the quantum memory at finite polarisations.

To calculate the electron--nuclear dynamics during the pulse sequence, we developed a numerically exact technique that maps the nuclear many-body state onto two one-dimensional chains of states, $\hat{\mathcal{S}}_\pm=\{\ket*{\hat{M}^{(k)}_\pm}|k=0,\cdots, N\}$. Here, the initial state $\ket*{\hat{M}_+^{(0)}}=\ket*{\hat{M}^{(0)}_-}=\ket{M}$ appears as the first link in both chains. The set $\hat{\mathcal{S}}_+$ ($\hat{\mathcal{S}}_-$) represents the set of states tied with the evolution of a positive (negative) spin wave.
The coupling structure of $\mathcal{H}_I$, taking the electron spin into account as well, is illustrated in Figure~\ref{fig:1-2}c, where a coupling rate of $\mathcal{G}_+$ between two neighbouring states is depicted with a double line and $\mathcal{G}_-$ with a single line.
This one-dimensional structure is largely attributed to the secular form of the interaction Hamiltonian, Eq.~\eqref{eq:6}: when the evolved state $U_I(t)\ket{\phi}\otimes\ket{M}$ is written out explicitly, all terms containing the products $S_+S_+$ or $S_-S_-$ vanish identically. In the Supplemental Material, we derive the form of the basis sets, $\hat{\mathcal{S}}_\pm$ and show that only neighbouring nuclear states are coupled by the spin wave operators entering $\mathcal{H}_I$,
\begin{align}
  \label{eq:3}
\begin{split}
  \mel*{\hat{M}_+^{(2n\pm 1)}}{\Phi_\zeta^+}{\hat{M}_+^{(2n)}} &= \mel*{\hat{M}_-^{(2n)}}{\Phi_\zeta^+}{\hat{M}_-^{(2n \mp 1)}}
=\mathcal{G}_\pm/\mathcal{A}_\zeta' 
\\
  \mel*{\hat{M}_+^{(2n)}}{\Phi^-_\zeta}{\hat{M}_+^{(2n \pm 1)}} &= \mel*{\hat{M}_-^{(2n \mp 1)}}{\Phi_\zeta^-}{\hat{M}_-^{(2n)}}
=\mathcal{G}_\pm/\mathcal{A}_\zeta'.
\end{split}
\end{align}

The asymmetry in rates is parametrised by a leakage factor, $\mathcal{G}_-/\mathcal{G}_+$, which indicates the extent to which the nuclear phase space is explored in the evolution. In the fully polarised case, where the initial state is $\ket*{M}=\ket*{\mathbf{0}}$ (and $\ket*{\hat{M}_+^{(1)}}=\ket{\mathbf{1}}$), we find $\mathcal{G}_+=g_\zeta,\; \mathcal{G}_-=0$ (no leakage), meaning that the three states $\ket*{\uparrow}\otimes\ket{M},\ket{\downarrow}\otimes\ket*{\hat{M}^{(1)}_+},\ket*{\downarrow}\otimes\ket*{M}$ are identically decoupled from the residual part of the two chains (which in Figure~\ref{fig:1-2}c is signified by $\mathcal{R}$), recovering the ideal state transfer dynamics of Eq.~\eqref{eq:ideal-dyn}. 
For finite initial nuclear polarisation, we generally have $\mathcal{G}_+<g_\zeta$ and $\mathcal{G}_->0$ (finite leakage), and thus the three states couple to the residual chains, $\mathcal{R}$. Figure~\ref{fig:1-2}d-e shows the electron and nuclear dynamics as the leakage factor $\mathcal{G}_-/\mathcal{G}_+$ is gradually changed from the ideal case of 0 (solid lines) to $30\%$ in linear steps (decreasing opacity). As $\mathcal{G}_-$ is increased, the system is more rapidly delocalised along the chain, leading to unwanted and uncontrollable electron--nuclear correlations that can be seen as damped oscillations in the nuclear and electronic populations.
When calculating the dynamics, it is necessary to truncate the nuclear chains of states to a certain $k$-index, $k^*$. As long as the occupation of the states $\ket*{\hat{M}_\pm^{(k^*)}}$ is appreciably small, the truncation remains a valid approximation. The necessary value of $k^*$ needed for convergence depends on the evolution time and the leakage factor $\mathcal{G}_-/\mathcal{G}_+$, which together determine how far into $\mathcal{R}$ the state will diffuse.

\begin{figure}
  \centering
  \includegraphics[width=\columnwidth]{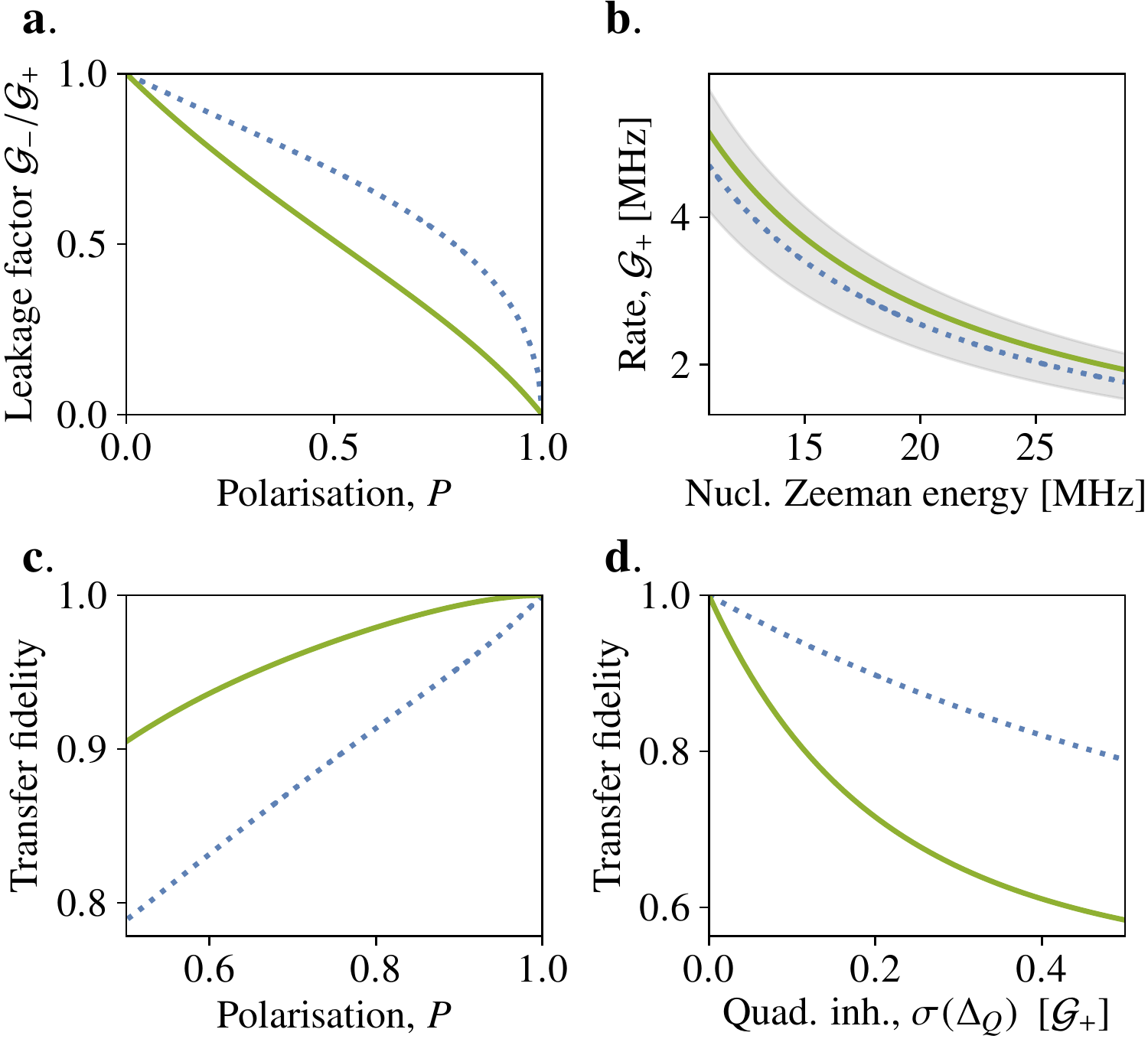}
  \caption{
{\bf a.} Leakage factor as ratio of coupling rates, $\mathcal{G}_-/\mathcal{G}_+$ as a function of nuclear polarisation. Green solid (blue dotted) lines denote $\zeta=2$ ($\zeta=1$). 
{\bf b.} Coupling rate $\mathcal{G}_+$ for $\zeta=1$ (blue dotted) and $\zeta=2$  (green solid) at $P=0.5$ for varying nuclear Zeeman splitting. The shaded area corresponds to the range of values from $P=0$ to $P=1$.
{\bf c.} Transfer fidelity of total write-in and read-out cycle as a function of polarisation. 
{\bf d.} Transfer fidelity at full polarisation ($P=1$) of total write-in and read-out cycle as a function of inhomogeneity in quadrupolar energy shift.}
  \label{fig:2}
\end{figure}

Figure~\ref{fig:2}a shows the ratio $\mathcal{G}_-/\mathcal{G}_+$ as a function of nuclear polarisation, averaged over the nuclear initial state distribution, $p(M)$, which we have taken thermal, with $I=3/2$ and $N \simeq 5\times 10^4$ (see Figure~\ref{fig:2}a). The relative ensemble standard deviations (not shown) are negligible, $\sim 10^{-3}$. An important conclusion drawn from Figure~\ref{fig:2}a is that the $\zeta=2$ mode is more robust to imperfect polarisation, owing to the different dependence of the leakage factor $\mathcal{G}_-/\mathcal{G}_+$ on polarisation for $\zeta=1$ and $\zeta=2$. 
The $\Phi^-_2$-transition becomes dark when the levels $m=+3/2,\; m=+1/2$ are depleted, whereas for $\Phi_1^-$, the level $m=-1/2$ needs to be depleted for this to happen. Indeed, the lower levels of a single spin manifold (Figure~\ref{fig:1-1}b) will be populated first, as the polarisation is decreased, thus enabling the $\Phi^-_1$ transition before the $\Phi_2^-$ transition. 

Figure~\ref{fig:2}b shows the calculated values of $\mathcal{G}_+$ as a function of nuclear Zeeman splitting, where the lines indicate a polarisation of 50\% and the shaded area shows the variation as the polarisation is sweeped from unity (maximum values) to 0 (minimum values). The coupling rate $\mathcal{G}_+$ is proportional to $\sin(2\theta)$ for the $\zeta=1$ mode and to $\sin^2\theta$ for $\zeta=2$. These angular prefactors have been assumed to be unity to maximinise the rates shown in Figure~\ref{fig:2}b, which can be converted to rates for any quadrupolar angle by multiplying them with these prefactors.

\paragraph{State transfer fidelity \textemdash}
 The state transfer fidelity is defined as the overlap between the initial electron state and the electron state after a full write--read cycle. We initialise the system in the state $\ket{\psi(0)}=\ket{\phi}\otimes\ket{M}$ and let the system evolve under the pulse sequence for a time $t_1$ to write the electron state into the nuclei, thus generating the state $\ket*{\psi(t_1)}$. After this, we trace out the electron to obtain the reduced nuclear state $\rho_\mathrm{n}(t_1)=\Tr_\mathrm{e}[\dyad*{\psi(t_1)}{\psi(t_1)}]$. To read the nuclear state back into the electron spin, we re-initialise the density operator in the state $\rho(t_1;0)=\dyad*{\downarrow}\otimes\rho_n(t_1)$ and let the system evolve under the pulse sequence for a time $t_2$, where the total density operator is $\rho(t_1;t_2)$. We then evaluate the overlap of the electron spin state (and tracing out the nuclei) with respect to the input state to assess the fidelity, $\mathcal{F}(t_1,t_2)=\Tr_\mathrm{n}[\mel{\phi'}{\rho(t_1;t_2)}{\phi'}]$, where $\ket{\phi'}=\alpha\ket{\uparrow}-\beta\ket{\downarrow}$. The fidelity, averaged over the six states $(\alpha,\beta)=(1,0), (0,1),\: \frac{1}{\sqrt{2}}(1,\pm 1),\: \frac{1}{\sqrt{2}}(1,\pm i)$, is presented in Figure~\ref{fig:2}c, for optimised values of $t_1$ and $t_2$. In the fully polarised case, the optimal $t_1$ and $t_2$ are simply $\pi/(2\mathcal{G}_+)$, but as the polarisation is decreased, coupling to $\mathcal{R}$ renormalises the effective coupling rate and therefore necessitates slightly (< 20\%) longer transfer times. As predicted, Figure~\ref{fig:2}c shows that the state transfer fidelity follows the polarisation dependence of the leakage factor, $\mathcal{G}_-/\mathcal{G}_+$, and accordingly that the fidelity is generally higher for the $\zeta=2$ mode if the polarisation is finite. In particular, for $\zeta=2$, the fidelity remains above 90\% throughout the polarisation range $50-100\%$.

\paragraph{Adjusting to quadrupolar energy shifts \textemdash}
The $I_z^j$-commuting contribution to the quadrupolar Hamiltonian can be written as $H_\mathrm{Q}^0=\sum_j\Delta_\mathrm{Q} (I_z^j)^2$, with $\Delta_\mathrm{Q}=B_\mathrm{Q}(\frac{1}{2}\sin^2\theta-\cos^2\theta)$. In general, $\Delta_\mathrm{Q}$ varies over the ensemble, and the individual spin components in the spin wave $\ket{\mathbf{1}}$ evolve with a phase factor $e^{-i\Delta_\mathrm{Q}^jt}$, building up a relative phase among the components on a time scale set by the ensemble variation of $\Delta_\mathrm{Q}^j$, denoted by $\sigma(\Delta_\mathrm{Q})$. As a result, $\ket{\mathbf{1}}$ rotates into a dark subspace, $\{\ket*{1_p}\}$, such that $\Phi_-\ket*{1_p}=0$ with a rate of $\gamma=1/(\zeta^2\sigma(\Delta_\mathrm{Q}))$~\cite{suppmat}. In Figure~\ref{fig:2}d, we show how the transfer fidelity at full nuclear polarisation depends on this inhomogeneity. As indicated, the energy scale of the inhomogeneity, $\sigma(\Delta_\mathrm{Q})$ must be compared to the coupling rate, $\mathcal{G}_+$, to assess its impact. Thus, with a realistic value of $\mathcal{G}_+$ in the $\mathrm{MHz}$ range, a quadrupolar inhomogeneity below $\sim 100\mathrm{\; kHz}$ does not degrade the transfer fidelity appreciably. 
Importantly, decoherence due to rotation into the dark subspace is only of concern during the transfer process: after the state has been transferred, the quadrupolar precession can be cancelled out by refocusing the $\{m=-I, m=-I+\zeta\}$ subspace using an NMR echo pulse~\cite{chekhovich2015suppression}.
In the case of a non-zero mean value of $\Delta_\mathrm{Q}$, the $m=-I$ to $m=-I+\zeta$ transition is shifted by $\delta=(\zeta^2-2I\zeta)\Delta_\mathrm{Q}$. The memory transfer is then simply effectuated by setting the pulse time delay to $\tau=\ell\pi/(\zeta\omega_\mathrm{Z}^\mathrm{n}+\delta)$.


During storage, we expect the dominant nuclear dephasing mechanism that determines the coherence time of the memory to be the electron--mediated nuclear dipole-dipole interaction, which scales inversely with the electron Zeeman splitting~\cite{wust2016role}. In the presence of this dephasing mechanism, the coherence time of the nuclear memory is tens of microseconds. If, however, the electron is removed from the quantum dot after its state is transferred to the nuclei, the only dephasing mechanism is the intrinsic neighbour dipole-dipole interaction, and the coherence time can be well into the millisecond regime~\cite{chekhovich2017decoherence}.
Importantly, the nuclei should be polarised as to increase the electron Zeeman splitting. This way, polarisation of the nuclei leads not only to increased fidelity in the transfer process, but also to further prolonged coherence time of the collective nuclear state.

\paragraph{Conclusion \textemdash}
For realistic experimental parameters, we have found that state transfer fidelities for a full read-write cycle as high as 90\% can be reached with a modest nuclear polarisation of 50\%. 
In addition, the theoretical and experimental techniques we have presented open new possibilities for further exploration and manipulation of the collective nuclear degrees of freedom, for example the generation of nuclear cat states, squeezed states and condensates.

\paragraph{Acknowledgements \textemdash}
We thank E. Chekhovich for helpful discussions. This work was supported by the ERC PHOENICS grant (617985) and the EPSRC Quantum Technology Hub NQIT (EP/M013243/1). D.A.G. acknowledges support from St John’s College Title A Fellowship.
E.V.D. and J.M. acknowledge funding from the Danish Council for Independent Research (Grant No. DFF- 4181-00416). C.L.G. acknowledges support from a Royal Society Dorothy Hodgkin Fellowship.

\clearpage
\widetext
\setcounter{page}{1}
\setcounter{figure}{0}
\setcounter{equation}{0}
\renewcommand{\theequation}{S\arabic{equation}}
\renewcommand{\thefigure}{S\arabic{figure}}
\renewcommand{\bibnumfmt}[1]{[S#1]}
\renewcommand{\citenumfont}[1]{S#1}
\begin{center}
\textbf{\Large Supplemental Material}\\
Here we elaborate on the expressions and results presented in the main text.
\end{center}

\section{Effective low-energy Hamiltonian for electron and nuclei in the presence of lattice strain}
\label{sec:hamilt-electr-nucl}
As the starting point of the analysis, we shall consider a singly charged quantum dot subject to a magnetic field of strength $B$, perpendicular to the growth axis. We define the $z$-direction as the direction of the field and denote the Cartesian electron spin operators in this reference frame by $S_\alpha,\; \alpha=x,y,z$. Along with the electron, the quantum dot contains $N$ nuclei, each with the spin operators $I_\alpha^j$, $\alpha=x,y,z;\; j=1,\cdots,N$. In the presence of uniaxial material strain, the quadrupole moment of the nuclear spins will couple to the electric field gradient of the strained lattice, as described by the Hamiltonian~\cite{SIbulutay2012quadrupolar,SIurbaszek2013nuclear}
\begin{align}
  \label{eq:S2}
 H_\mathrm{Q}=\sum_j B_\mathrm{Q}^j\qty[(I_x^{j})^2\sin^2\theta^j + \frac{1}{2}(I_x^jI_z^j+I_z^jI_x^j)\sin2\theta^j + (I_z^{j})^2\cos^2\theta^j ],
\end{align}
where $\theta$ is the angle between the quadrupolar axis and the magnetic field and $B_\mathrm{Q}$ is the strength of the quadrupolar interaction.
Furthermore, the electron and nuclei interact via the hyperfine interaction, $H_\mathrm{hf}=\sum_j2A^j[S_zI_z^j + S_xI_x^j+S_yI_y^j]$. For appreciable external magnetic field strength, the last two terms in $H_\mathrm{hf}$ describe electron--nucleus flip-flop processes that are perturbatively suppressed a factor of $A^j/(\omega_Z^\mathrm{e}-\omega_Z^\mathrm{n})\ll 1$, for which reason these terms are typically neglected. In our case, this is well justified by the fact that the leading order perturbative processes governing the electron--nuclear energy exchange occur at a much higher rate, as we shall see. For these reasons, we take $H_\mathrm{hf}=\sum_j 2A^jS_zI_z^j$. The total Hamiltonian for the electron and nuclear bath is then 
\begin{align}
  \label{eq:S1}
  H=\omega_\mathrm{Z}^eS_z + \omega_\mathrm{Z}^\mathrm{n}\sum_j I_z^j + H_\mathrm{Q} + H_\mathrm{hf},
\end{align}
with $\omega_\mathrm{Z}^\mathrm{e}$ and $\omega_\mathrm{Z}^\mathrm{n}$ the electron and nuclear Zeeman energies, respectively.

From the quadrupolar interaction, $H_\mathrm{Q}$, we now extract the contribution that commutes with $I_z^j$ and thus does not couple different nuclear Zeeman eigenstates, which we denote by $H_\mathrm{Q}^0$. The remainder, $V_\mathrm{Q}:=H_\mathrm{Q}-H_\mathrm{Q}^0$ is then entirely off-diagonal in the nuclear Zeeman eigenbasis. Specifically, we have~\cite{SIgangloff2019magnon}
\begin{align}
  \label{eq:S6}
\begin{split}
  H_\mathrm{Q}^0&=\sum_j B_\mathrm{Q}^j\qty{\frac{1}{2}\qty[(I_x^j)^2+(I_y^j)^2]\sin^2\theta^j + (I_z^j)^2\cos^2\theta^j} \\
  V_\mathrm{Q}  &=\sum_j \frac{1}{2} B_\mathrm{Q}^j\qty{[(I_x^j)^2-(I_y)^2]\sin^2\theta^j+[I_x^jI_z^j+I_z^jI_x^j]\sin2\theta^j}
\end{split}
\end{align}
In strong field conditions where $\omega_Z^\mathrm{n}\gg B_\mathrm{Q}^j$, transitions between nuclear Zeeman eigenstates caused by $V_\mathrm{Q}$ are not energetically allowed to first order. To eliminate $V_\mathrm{Q}$ from $\hat{H}$ and replace it with the appropriate corrections describing energetically allowed processes, we use a Schrieffer-Wolff transformation with generator
\begin{align}
  \label{eq:S7} F=i\sum_j\frac{B_\mathrm{Q}^j}{2\omega_Z^\mathrm{n}}\qty{\frac{1}{2}(I_x^jI_y^j+I_y^jI_x^j)\sin^2\theta^j+(I_y^jI_z^j+I_z^jI_y^j)\sin2\theta^j}.
\end{align}
Up to second order in the perturbation parameters $B_\mathrm{Q}^j/\omega_Z^\mathrm{n}$ and $A^j/\omega_Z^\mathrm{n}$, we find the transformed Hamiltonian $H'=e^{F}\hat{H}e^{-F}\simeq\hat{H}_\mathrm{e}+H_\mathrm{n}'+H_\mathrm{hf}+V_\mathrm{Q}'$, where~\cite{SIgangloff2019magnon}
\begin{align}
  \label{eq:S8}
 H_\mathrm{n}'&=\omega_Z^\mathrm{n}\sum_jI_z^j +H_\mathrm{Q}^0+[F,V_\mathrm{Q}], \\
  V_\mathrm{Q}'-&=-S_z\sum_j\frac{A^jB_\mathrm{Q}^j}{\omega_Z^\mathrm{n}}\qty{[(I_x^j)^2-(I_y^2)^j]\sin^2\theta^j + [I_x^jI_z^j+I_z^jI_x^j]\sin2\theta^j}.
\end{align}
Note that $H_\mathrm{n}'$ commutes with $I_z^j$ and only leads to an anharmonic energy shift of the single-nucleus spin ladders, such that the Zeeman eigenstates of the $j$'th nucleus, $\ket{m}_j$, have the energies $m\omega_Z^\mathrm{n}+m^2\Delta_\mathrm{Q}^j$, where $\Delta_\mathrm{Q}^j=B_\mathrm{Q}^j\qty(\frac{1}{2}\sin^2\theta^j-\cos^2\theta^j)$. In contrast, $V_\mathrm{Q}'$ describes a quadrupolar dressing of the hyperfine interaction that generates a noncollinear collective interaction between the electronic and nuclear spins.

\subsection{Hyperfine coupling distribution}
\label{sec:hyperf-coupl-distr}
The hyperfine coupling distribution, $A^j$, is highly non-uniform due to the inhomogeneous form of the electron wavefunction. For all practical purposes when calculating properties of the system, we obtain the hyperfine distribution numerically by taking the electron density Gaussian,
\begin{align}
  \label{eq:S35}
  \rho_e(\mathbf{r})=\prod_{\alpha=x,y,z}\frac{e^{-r_\alpha^2/(2L_\alpha^2)}}{\sqrt{2\pi L_\alpha^2}},
\end{align}
and evaluating $\rho_e$ in the points of a cubic lattice of size $L_x\times L_y\times L_z$. We have taken parameters for arsenic nuclei in GaAs and a quantum dot with $L_x=L_y=10\mathrm{\; nm},\; L_z=1\mathrm{\; nm}$, consistent with e.g. Ref.~\cite{SIhuo2014light}.

\section{Pulse sequence}
\label{sec:pulse-sequence}
\begin{figure}
  \centering
  \includegraphics[width=0.5\textwidth]{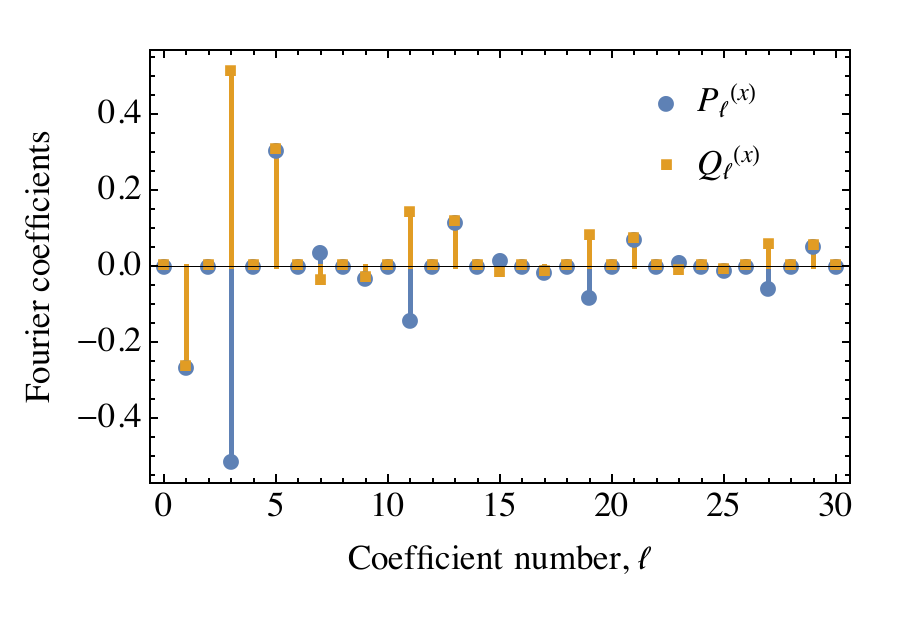}
  \caption{Fourier coefficients for the pulse modulation function $h_x$}
  \label{fig:fourier-coefficients}
\end{figure}

By rotating the electron spin with a series of short pulses in conjunction with the free evolution of the system, it is possible to engineer the dynamics of the electron--nuclear system by enhancing and quenching various terms in the Hamiltonian controllably. Here, we adapt a pulse sequence developed for a central spin coupled a nuclear environment with a few, energetically separated spins~\cite{SIschwartz2018robust} to our case of a mesoscopic bath with an energetically dense spectrum. The pulse sequence is described in detail in Ref.~\cite{SIschwartz2018robust}, but for completeness we present the central features here before showing how the sequence acts on the system studied in this work. The pulse cycle of the sequence can be written as 
\begin{align}
  \label{eq:S9}
  y_{\pi/2} \;\frac{\tau/4}{\hspace{0.8cm}}\; x_{-\pi} \;\frac{\tau/4}{\hspace{0.8cm}}\; y_{\pi/2}\; x_{\pi/2}\;\frac{\tau/4}{\hspace{0.8cm}}\;y_{\pi}\;\frac{\tau/4}{\hspace{0.8cm}}\;x_{\pi/2},
\end{align}
where $q_\phi$ ($q=x,y$) denotes a fast coherent rotation of the electron spin corresponding to the unitary operation $e^{-i\phi S_q }$ and $\frac{\tau/4}{\hspace{0.8cm}}$ denotes free evolution of the system during the time interval $\tau/4$.
The unitary evolution operator for a cycle of the sequence can then be expressed as 
\begin{align}
  \label{eq:S10}
  U_\mathcal{C}=e^{-i\frac{\pi}{2}S_x}U_0\qty(\frac{\tau}{4})e^{-i\pi S_y}U_0\qty(\frac{\tau}{4})e^{-i\frac{\pi}{2}S_x}
              e^{-i\frac{\pi}{2}S_y}U_0\qty(\frac{\tau}{4})e^{+i\pi S_x}U_0\qty(\frac{\tau}{4})e^{-i\frac{\pi}{2}S_y},
\end{align}
where $U_0(t)=e^{-iH't}$ is the free evolution operator of the system. From Eq.~\eqref{eq:S10} it can be shown that the time evolution operator over two consecutive cycles can be written as the dynamics generated by a time dependent Hamiltonian, $\mathcal{H}(t)$, such that $U_\mathcal{C}^2=\mathcal{T} e^{-i\int_0^{2\tau}\dd s \mathcal{H}(s)}$, where $\mathcal{T}$ is the chronological time-ordering operator. The pulse sequence Hamiltonian, $\mathcal{H}(t)$, is obtained as $H'$ under the substitution $S_z\rightarrow h_x(t)S_x+h_y(t)S_y$, with the piecewise constant functions, which in the interval $t\in [0,2\tau]$ take the values
\begin{align}
  \label{eq:S13}
  h_x(t)=\begin{cases}
0, & t\in[0,\tau/2[ \\
-1, &  t\in[\tau/2, 3\tau/4[ \\ 
+1, & t\in [3\tau/4,\tau[ \\
0, & t\in[\tau,3\tau/2[ \\
+1, & t\in [3\tau/2,7\tau/4[ \\
-1, & t\in [7\tau/4, 2\tau[
\end{cases}
\end{align} 
with $h_y(t)=h_x(t+\tau/2)$. The time-dependent functions feature the periodic property $h_x(t+2\tau \ell)=h_x(t),\; h_y(t+2\tau \ell)=h_y(t)$, where $\ell=0,1,\cdots $. Due to this periodicity, we can write the $h_\alpha$-functions in terms of their discrete Fourier series as
\begin{align}
  \label{eq:S14}
  h_\alpha(t)=\sum_{\ell=0}^\infty P_\ell^{(\alpha)}\cos(\omega_\ell t) + Q_\ell^{(\alpha)}\sin(\omega_\ell t),
\end{align}
with $\omega_\ell=\pi\ell/\tau$. The Fourier components are calculated from $h_\alpha$ as 
\begin{align}
  \label{eq:S15}
\begin{split}
  P_\ell^{(\alpha)} &=\frac{1}{\tau}\int_0^{2\tau}h_\alpha(t)\cos(\omega_\ell t)\\
  Q_\ell^{(\alpha)} &=\frac{1}{\tau}\int_0^{2\tau}h_\alpha(t)\sin(\omega_\ell t),
\end{split}
\end{align}
which due to the relation between $h_x$ and $h_y$ satisfy $P_\ell^{(x)}=P_\ell^{(y)}, \; Q_\ell^{(x)}=-Q_\ell^{(y)}$. Furthermore, the coefficients are only nonzero for odd $\ell$. These coefficients are plotted in Fig.~\ref{fig:fourier-coefficients}, which shows that the $\ell=3$ coefficients are largest. For these maximal coefficients, we have
\begin{align}
  \label{eq:S12}
  P^{(x)}_3=-Q^{(x)}_3=-\frac{2+2\sqrt{2}}{3\pi} \simeq -0.51 
\end{align}

We now assume that the quadrupolar energy shift contained in $H_\mathrm{n}'$ is negligible compared to $\omega_Z^\mathrm{n}$, which can be ensured if the angle of the strain axis relative to the magnetic field is close to $\theta_0=\arctan(1/\sqrt{2})$. We can then write the pulse sequence Hamiltonian as $\mathcal{H}(t)\simeq\mathcal{H}_0+\mathcal{H}_I(t)$,
\begin{align}
  \label{eq:S16}
\mathcal{H}_0=\omega_Z^\mathrm{n}\sum_jI_z^j, \;\;
  \mathcal{H}_I(t)=-[h_x(t)S_x+h_y(t)S_y]\qty{\sum_jA^jI^j_z+\mathcal{A}_1(\Phi_1^++\Phi_1^-) + \mathcal{A}_2(\Phi_2^++\Phi_2^-)},
\end{align}
where $\Phi_\zeta^{+}=(\Phi_\zeta^{-})^\dagger$ and
\begin{align}
  \label{eq:S34}
  &\Phi_1^{-}:= \sum_j a_{1,j} \qty(I_-^jI_z^j + I_z^jI_-) , &\mathcal{A}_1:=\frac{1}{2}\sum_j\frac{A^jB_\mathrm{Q}^j\sin2\theta^j}{\omega_Z^\mathrm{n}}, \;\; \;\; \;\;\;\;
&a_{1,j}:=\frac{1}{\mathcal{A}_1}\frac{A^jB_\mathrm{Q}^j\sin2\theta^j}{\omega_Z^\mathrm{n}},\\ 
  &\Phi_2^{-}:= \sum_j a_{2,j} (I_-^{j})^2,  &\mathcal{A}_2:=\frac{1}{2}\sum_j\frac{A^jB_\mathrm{Q}^j\sin^2\theta^j}{\omega_Z^\mathrm{n}},\;\;\;\; \;\; \;\;
&a_{2,j}:=\frac{1}{\mathcal{A}_2}\frac{A^jB_\mathrm{Q}^j\sin^2\theta^j}{\omega_Z^\mathrm{n}},
\end{align}
such that $\sum_ja_{\zeta,j}=1$. Note that in the free evolution between the Raman pulses, the drives are absent and the rotating frame can be defined with respect to an arbitrary frequency, which we set to $\omega_Z^\mathrm{e}$, leading to $\Delta=0$. Moving into the interaction picture set by $\mathcal{H}_0$, the interaction Hamiltonian is
\begin{align}
  \label{eq:S3}
  \hat{\mathcal{H}}_I(t)=e^{i\mathcal{H}_0t}\mathcal{H}_I(t)e^{-i\mathcal{H}_0t}=-[h_x(t)S_x+h_y(t)S_y]\qty{\sum_jA^jI^j_z+\mathcal{A}_1(e^{+i\omega_Z^\mathrm{n} t}\Phi_1^++e^{-i\omega_Z^\mathrm{n} t}\Phi_1^-) + \mathcal{A}_2(e^{+2i\omega_Z^\mathrm{n} t}\Phi_2^++e^{-2i\omega_Z^\mathrm{n} t}\Phi_2^-)}.
\end{align}
At this point, we express $h_\alpha(t)$ in terms of their Fourier expansions,
\begin{align}
  \label{eq:S17}
\begin{split}
  \hat{\mathcal{H}}_I(t)&= -\sum_{\ell=1}^\infty\qty{\qty[P_\ell^{(x)}\cos(\omega_\ell t)+Q_\ell^{(x)}\sin(\omega_\ell t)]S_x+\qty[P_\ell^{(x)}\cos(\omega_\ell t)-Q_\ell^{(x)}\sin(\omega_\ell t)]S_y]}\\ &\hspace{0.1\columnwidth}\times \qty{\sum_jA^jI^j_z+\mathcal{A}_1(e^{+i\omega_Z^\mathrm{n} t}\Phi_1^++e^{-i\omega_Z^\mathrm{n} t}\Phi_1^-) + \mathcal{A}_2(e^{+2i\omega_Z^\mathrm{n} t}\Phi_2^++e^{-2i\omega_Z^\mathrm{n} t}\Phi_2^-)}.
\end{split}
\end{align}
Furthermore, we assume that the delay time of the pulse sequence, $\tau$, is chosen such that one of the of the discrete Fourier frequencies, $\omega_\ell$, is close to either $\omega_Z^\mathrm{n}$ or $2\omega_Z^\mathrm{n}$. We shall denote this particular $\ell$-index by $\ell^*$ and say that $\omega_{\ell^*}=\zeta \omega_Z^\mathrm{n}$, where $\zeta$ is either 1 or 2. This resonance of the pulse sequence is obtained by setting $\tau=\pi\ell^*/(\zeta\omega_Z^\mathrm{n})$. As a result, the Fourier frequencies, $\omega_\ell$, are separated by $\Delta\omega=\zeta\omega_Z^\mathrm{n}/\ell^*$. Provided that $\ell^*$ is of order unity, all terms in $\hat{\mathcal{H}}_I(t)$ that are not resonant with $\zeta\omega_Z^\mathrm{n}$ will average to zero in the dynamical evolution of the system. In particular, there does not exist a pair of odd indices $(\ell_1,\ell_2)$ such that $\ell_1\Delta\omega=\omega_Z^\mathrm{n}$ and $\ell_2\Delta\omega=2\omega_Z^\mathrm{n}$. Since the Fourier coefficients $P_\ell^{(\alpha)}$ and $Q_\ell^{(\alpha)}$ are only nonzero for odd $\ell$, this means that the pulse sequence will never be resonant with both collective nuclear transitions simultaneously. Removing all rapidly rotating terms from Eq. \eqref{eq:S17} leaves us with
\begin{align}
  \label{eq:S19}
  \hat{\mathcal{H}}_I(t)\simeq -\mathcal{A}_\zeta\frac{1}{2}\Phi_\zeta^+\qty[P_{\ell^*}^{(x)}(S_x+S_y)+iQ_{\ell^*}^{(x)}(S_x-S_y)] 
+ \mathcal{A}_\zeta\frac{1}{2}\Phi_\zeta^-\qty[P_{\ell^*}^{(x)}(S_x+S_y)-iQ_{\ell^*}^{(x)}(S_x-S_y)].
\end{align}
At this point, we choose $\ell^*=3$ which we found to have the largest Fourier component. Further, we rotate the electron spin coordinates by an angle of $\pi/4$ in the $xy$-plane, such that $S_x\rightarrow (S_x+S_y)/\sqrt{2}, \; S_y\rightarrow (S_x-S_y)/\sqrt{2}$, leading to
\begin{align}
  \label{eq:S20}
  \hat{\mathcal{H}}_I\simeq \frac{\sqrt{2}+2}{3\pi}\mathcal{A}_\zeta (\Phi_\zeta^+S_-+\Phi_\zeta^-S_+).
\end{align}

\section{Nuclear chain of states under pulse sequence dynamics}
\label{sec:dynamics}
The pulse sequence interaction Hamiltonian, $\hat{\mathcal{H}}_I$, in Eq.~\eqref{eq:S20} describes the interaction between the electron and the nuclear spin bath. Due to its secular form, many of the terms in the expansion of the time evolution operator, $\mathcal{U}$, become zero, because $S_+^2=S_-^2=0$. As a result, we can write $\mathcal{U}$ as 
\begin{align}
  \label{eq:S18}
\begin{split}
  \mathcal{U}(t)&=e^{-i\hat{\mathcal{H}}_I t}=\mathbb{I}+\qty(\frac{-i\mathcal{A}_\zeta t}{2})^2(\Phi_\zeta^+S_-+\Phi_\zeta^-S_+)
+ \frac{1}{2!}\qty(\frac{-i\mathcal{A}_\zeta t}{2})^2(\Phi_\zeta^-\Phi_\zeta^+ S_+S_- + \Phi_\zeta^+\Phi_\zeta^-S_-S_+)
\\ &\hspace{0.1\columnwidth}+ \frac{1}{3!}\qty(\frac{-i\mathcal{A}_\zeta t}{2})^3(\Phi_\zeta^+\Phi_\zeta^-\Phi_\zeta^+ S_-S_+S_- + \Phi_\zeta^-\Phi_\zeta^+\Phi_\zeta^-S_+S_-S_+)+\cdots
\\ &=\sum_{k=0}^\infty \frac{1}{(2k)!}\qty(\frac{-i\mathcal{A}_\zeta t}{2})^{2k} \qty{(\Phi_\zeta^+\Phi_\zeta^-)^k(S_-S_+)^k+(\Phi_\zeta^-\Phi_\zeta^+)^k(S_+S_-)^k}
\\ &\hspace{0.1\columnwidth} +\sum_{k=0}^\infty \frac{1}{(2k+1)!}\qty(\frac{-i\mathcal{A}_\zeta t}{2})^{2k+1} \qty{\Phi_\zeta^-(\Phi_\zeta^+\Phi_\zeta^-)^kS_+(S_-S_+)^k+\Phi_\zeta^+(\Phi_\zeta^-\Phi_\zeta^+)^kS_-(S_+S_-)^k}
\end{split}
\end{align}
At this point, we consider the initial nuclear state to be a classical mixture of product states. Denoting a general nuclear product state by $\ket{M}=\ket{m_1,\cdots,m_\mathrm{n}}$, the initial nuclear density operator can be written as $\rho_\mathrm{n}(0)=\sum_M p(M)\dyad{M}$, which does not need to be internuclear factorisable, but can generally contain classical correlations, i.e. generally $\rho_\mathrm{n}(0)\neq\bigotimes \rho_j$. For practical purposes, we can then calculate the dynamics for a given state $\ket{M}$ and subsequently average over the distribution $p(M)$. From Eq.~\eqref{eq:S18}, we see that with $\ket{M}$ as the initial nuclear state, the evolution of the system will gradually populate the states $\Phi_\zeta^\pm\ket{M},\; \Phi_\zeta^\mp \Phi_\zeta^\pm\ket{M}, \Phi_\zeta^\pm\Phi_\zeta^\mp \Phi_\zeta^\pm\ket{M}$ and so forth. Our strategy for calculating the dynamics of the electron--nuclear system is to orthonormalise the set
$$\mathcal{S}_\pm(k^*):=\{(\Phi_\zeta^\mp\Phi_\zeta^\pm)^k\ket{M}, \Phi_\zeta^\pm(\Phi_\zeta^\mp\Phi_\zeta^\pm)^k\ket{M}|k\leq k^*\},$$
containing states generated by the evolution $\mathcal{U}$ up to a certain truncation index, $k^*$. Note that by $\mathcal{S}_+(k^*)$ and $\mathcal{S}_-(k^*)$, we understand two distinct sets, which we wish to orthonormalise separately.
We shall define the normalised state $\ket*{M^{(k)}_\pm;\zeta}$ as 
\begin{align}
  \label{eq:S22}
  \ket*{M^{(k)}_\pm;\zeta}=\begin{cases}
\frac{(\Phi_\zeta^\mp\Phi_\zeta^\pm)^k\ket{M}}{\sqrt{\mel{M}{(\Phi_\zeta^\mp\Phi_\zeta^\pm)^{2k} }{M}}}, & k \mathrm{\; even} \\ \\
\frac{\Phi_\zeta^\pm(\Phi_\zeta^\mp\Phi_\zeta^\pm)^{k-1}\ket{M}}{\sqrt{\mel{M}{(\Phi_\zeta^\mp\Phi_\zeta^\pm)^{2k-1}}{M}}}, & k \mathrm{\; odd},
\end{cases}
\end{align}
such that $\bar{\mathcal{S}}_\pm(k^*):=\{\ket*{M^{(k)}_\pm;\zeta}|k\leq k^*\}$ is simply the normalised form of $\mathcal{S}_\pm(k^*)$. The goal is now to perform Gram-Schmidt orthonormalisation to $\bar{\mathcal{S}}_\pm(k^*)$ in order to derive the orthonormal set $\hat{\mathcal{S}}_\pm(k^*)$, whose elements we shall denote by $\ket*{\hat{M}^{(k)}_\pm;\zeta}$. First, we see that $\ket*{M}=\ket*{M^{(0)}_\pm;\zeta}$ and define $\ket*{\hat{M}^{(0)}_\pm;\zeta}:=\ket{M}$. The Gram-Schmidt orthonormalisation strategy now gives the recursive relation
\begin{align}
  \label{eq:S23} \ket*{\hat{M}^{(1)}_\pm;\zeta}&=\frac{(\mathbb{I}-\dyad*{M^{(0)}_\pm;\zeta})\ket*{M^{(1)}_\pm;\zeta}}{\sqrt{\mel*{M^{(1)}_\pm;\zeta}{(\mathbb{I}-\dyad*{M^{(0)}_\pm;\zeta}) }{M^{(1)}_\pm;\zeta}}}, \\ 
\ket*{\hat{M}^{(k)}_\pm;\zeta}&=\frac{\qty(\mathbb{I}-\sum_{l=0}^{k-1}\dyad*{\hat{M}^{(l)}_\pm;\zeta})\ket*{M^{(k)}_\pm;\zeta}}{\sqrt{\mel*{M^{(k)}_\pm;\zeta}{\qty(\mathbb{I}-\sum_{l=0}^{k-1}\dyad*{\hat{M}^{(l)}_\pm;\zeta}) }{M^{(k)}_\pm;\zeta}}}.
\end{align}
The structure of the orthonormal set $\hat{\mathcal{S}}(k^*)$ becomes clear when writing the states out explicitly. To do so, we first introduce the convenient shorthand notation
\begin{align}
\label{eq:S28}
\ket{M;(\Delta,j),(\Delta',{j'}),\cdots}:=\ket{m_1,\cdots,(m_j+\Delta),\cdots,(m_{j'}+\Delta'),\cdots,m_\mathrm{n}},
\end{align}
and the prefactors (not to be confused with the Fourier coefficients in Sec.~\ref{sec:pulse-sequence})
\begin{align}
  \label{eq:S27}
\begin{split}
  P_\pm^{(1)}(m)&=(2m\pm 1)\sqrt{I(I+1)-m(m\pm 1)} \\
  P_\pm^{(2)}(m)&=\sqrt{I(I+1)-m(m\pm1)}\sqrt{I(I+1)-(m\pm1)(m\pm2)},
\end{split}  
\end{align}
such that $\Phi_\zeta^\pm\ket{M}=\sum_j a_{\zeta,j}P_\pm^{(\zeta)}(m_j)\ket{M,(\pm\zeta,j)}$. Note that the prefactor $P^{(\zeta)}_\pm(m)$ automatically becomes zero if the transition $m\rightarrow m\pm\zeta$ is not allowed. The first four states in $\hat{\mathcal{S}}_+(k^*)$ and $\hat{\mathcal{S}}_-(k^*)$ can then be written as 
\begin{align}
  \label{eq:S26}
\begin{split}
\ket*{M^{(0)}_\pm;\zeta}&=\ket{M}\\
  \ket*{\hat{M}^{(1)}_\pm;\zeta}&=\frac{1}{\mathcal{N}^{(1)}_\pm(M;\zeta)}\sum_ja_{\zeta,j}P^{(\zeta)}_\pm(m_j)\ket{M;(\pm\zeta,j)}, \\
  \ket*{\hat{M}^{(2)}_+;\zeta}&=\ket*{\hat{M}^{(2)}_-;\zeta}=\frac{1}{\mathcal{N}^{(2)}(M;\zeta)}\sum_{\langle j_1j_2\rangle}a_{\zeta,{j_1}}a_{\zeta,{j_2}}P^{(\zeta)}_+(m_{j_1})P_-^{(\zeta)}(m_{j_2})\ket{M;(+\zeta,j_1),(-\zeta,j_2)}, \\
\ket*{\hat{M}^{(3)}_+;\zeta}&=\frac{1}{\mathcal{N}_+^{(3)}(M;\zeta)}
\Big\{
\sum_{\langle j_1 j_2j_3\rangle}a_{\zeta,{j_1}}a_{\zeta,{j_2}}a_{\zeta,{j_3}}
P^{(\zeta)}_+(m_{j_1})P_-^{(\zeta)}(m_{j_2})P_+^{(\zeta)}(m_{j_3})
\ket{M;(+\zeta,{j_1}),(-\zeta,{j_2}),(+ \zeta,{j_3})} 
\\ &\hspace{0.2\columnwidth}+ \sum_{\langle j_1 j_2\rangle} a_{j_1}^2a_{j_2}
P_+^{(\zeta)}(m_{j_1})P_+^{(\zeta)}(m_{j_1}+\zeta)P_-^{(\zeta)}(m_{j_2})
\ket{M;(+2\zeta,j_1),(-\zeta,j_2)}
\Big\},\\
\ket*{\hat{M}^{(3)}_-;\zeta}&=\frac{1}{\mathcal{N}^{(3)}_-(M;\zeta)}
\Big\{
\sum_{\langle j_1 j_2j_3\rangle}a_{\zeta,{j_1}}a_{\zeta,{j_2}}a_{\zeta,{j_3}}
P^{(\zeta)}_+(m_{j_1})P_-^{(\zeta)}(m_{j_2})P_+^{(\zeta)}(m_{j_3})
\ket{M;(+\zeta,{j_1}),(-\zeta,{j_2}),(- \zeta,{j_3})} 
\\ &\hspace{0.2\columnwidth}+ \sum_{\langle j_1 j_2\rangle} a_{j_1}a_{j_2}^2
P_+^{(\zeta)}(m_{j_1})P_-^{(\zeta)}(m_{j_2})P_-^{(\zeta)}(m_{j_2}-\zeta)
\ket{M;(+\zeta,j_1),(-2\zeta,j_2)}
\Big\},
\end{split}
\end{align}
where $\mathcal{N}_\pm^{(k)}(M;\zeta)$ is a normalisation factor and $\sum_{\langle j_1\cdots j_n\rangle}$ denotes a summation over the $n$ indices (each running from $1$ to $N$) that only includes terms where no pair of indices are equal. Extending this sequence of states to higher values of $k$ is a tedious, but straightforward task.
For the special case of $I=3/2$, the situation vastly simplifies due to the identity $P_\pm^{(\zeta)}(m)P_\pm^{(\zeta)}(m\pm 1)=0$, thus eliminating the possibility of multiple noncollinear excitations of the same nuclear spin. As a result, for $I=3/2$, we can write any state in $\hat{\mathcal{S}}(k^*)$ in a general form as 
\begin{align}
  \label{eq:S27}
\begin{split}
\ket*{\hat{M}^{(k)}_\pm;\zeta}&=\frac{\sum_{\langle j_1\cdots j_k\rangle} a_{\zeta,1}\cdots a_{\zeta,k}P_\pm^{(\zeta)}(m_{j_1})P_\mp^{(\zeta)}(m_{j_2})\cdots P_{\pm\lambda_k}^{(\zeta)}(m_{j_k})\ket*{M;(\pm\zeta,j_{1}),(\mp\zeta,j_2),\cdots,(\pm\lambda_k\zeta,j_k)}}{\sqrt{\sum_{\langle j_1\cdots j_k\rangle} [a_{\zeta,1}\cdots a_{\zeta,k}P_\pm^{(\zeta)}(m_{j_1})P_\mp^{(\zeta)}(m_{j_2})\cdots P_{\pm\lambda_k}^{(\zeta)}(m_{j_k})]^2}},
\end{split}
\end{align}
where $\lambda_k=(-1)^{k+1}$. In general, we find that for all even values of $k$, $\ket*{\hat{M}_+^{(k)};\zeta}=\ket*{\hat{M}_-^{(k)};\zeta}$, and we might thus drop the $\pm$-index on $\ket*{\hat{M}_\pm^{(k)};\zeta}$ for even $k$. 

For the purpose of calculating the dynamics, we are generally interested in knowing how the interaction Hamiltonian, $\hat{\mathcal{H}}_I$, couples the diffenent states in $\hat{\mathcal{S}}_\pm(k^*)$. The general structure of $\hat{\mathcal{S}}(k^*)$ as presented in Eqs.~\eqref{eq:S26} and \eqref{eq:S27}, leads to the selection rule
\begin{align}
  \label{eq:S32}
  \mel*{\hat{M}_\alpha^{(k)};\zeta}{\Phi_\zeta^\beta}{\hat{M}_\gamma^{(k')};\zeta}=0\;\mathrm{if}\; \abs{k-k'} \neq 1,
\end{align}
meaning that we only need to evaluate transition matrix elements between neighbouring states in $\hat{\mathcal{S}}_\pm(k^*)$. For general $I$, we find the elements between the states in Eq.~\eqref{eq:S26}
\begin{align}
  \label{eq:S33}
\begin{split}
  \mel*{\hat{M}_\pm^{(1)};\zeta}{\Phi_\zeta^\pm}{M^{(0)};\zeta}&=\sqrt{\sum_j [a_{\zeta,j}P_\pm(m_j)]^2} =:\Omega_\pm(M;\zeta), \\
\mel*{\hat{M}_\pm^{(1)};\zeta}{\Phi_\zeta^\mp}{M^{(0)};\zeta}&=0, \\
\mel*{\hat{M}^{(2)};\zeta}{\Phi_\zeta^\pm}{\hat{M}_\pm^{(1)};\zeta}&=0, \\
\mel*{\hat{M}^{(2)};\zeta}{\Phi_\zeta^\mp}{\hat{M}_\pm^{(1)};\zeta}&=\frac{1}{\Omega_\pm(M;\zeta)}\sqrt{\sum_{\langle j_1j_2\rangle }[a_{\zeta,j_1}a_{\zeta,j_2}
P_+(m_{j_1})P_-(m_{j_2})]^2}, \\
\mel*{\hat{M}^{(3)}_\pm;\zeta}{\Phi_\zeta^\mp}{\hat{M}^{(2)};\zeta}&=0, \\
\mel*{\hat{M}^{(3)}_+;\zeta}{\Phi_\zeta^+}{\hat{M}^{(2)};\zeta} &=\Big\{\sum_{\langle j_1j_2\rangle }[a_{\zeta,j_1}a_{\zeta,j_2}
P_+(m_{j_1})P_-(m_{j_2})]^2\Big\}^{-1/2}\Big\{\sum_{\langle j_1 j_2j_3\rangle}[a_{\zeta,{j_1}}a_{\zeta,{j_2}}a_{\zeta,{j_3}}
P^{(\zeta)}_+(m_{j_1})P_-^{(\zeta)}(m_{j_2})P_+^{(\zeta)}(m_{j_3})]^2
\\ &\hspace{0.1\columnwidth} + \sum_{\langle j_1 j_2\rangle} [a_{j_1}^2a_{j_2}
P_+^{(\zeta)}(m_{j_1})P_+^{(\zeta)}(m_{j_1}+\zeta)P_-^{(\zeta)}(m_{j_2})]^2\Big\} \\
\mel*{\hat{M}^{(3)}_-;\zeta}{\Phi_\zeta^-}{\hat{M}^{(2)};\zeta} &=\Big\{\sum_{\langle j_1j_2\rangle }[a_{\zeta,j_1}a_{\zeta,j_2}
P_+(m_{j_1})P_-(m_{j_2})]^2\Big\}^{-1/2}\Big\{\sum_{\langle j_1 j_2j_3\rangle}[a_{\zeta,{j_1}}a_{\zeta,{j_2}}a_{\zeta,{j_3}}
P^{(\zeta)}_+(m_{j_1})P_-^{(\zeta)}(m_{j_2})P_-^{(\zeta)}(m_{j_3})]^2
\\ &\hspace{0.1\columnwidth} + \sum_{\langle j_1 j_2\rangle} [a_{j_1}^2a_{j_2}
P_-^{(\zeta)}(m_{j_1})P_-^{(\zeta)}(m_{j_1}-\zeta)P_+^{(\zeta)}(m_{j_2})]^2\Big\}.
\end{split}
\end{align}
Because $\Phi^-=(\Phi^+)^\dagger$, we only need to find the matrix elements for one of the two operators, since the elements of the other will be given thereby. Here we find the elements of $\Phi^+$.

For $N\gg 1$ and $k\ll N$, we find the approximation
\begin{align}
  \label{eq:S29}
\begin{split}
  \mel*{\hat{M}_\pm^{(k'+2)}}{\Phi^+}{\hat{M}_\pm^{k+2}} &\simeq \mel*{\hat{M}_\pm^{(k')}}{\Phi^+}{\hat{M}_\pm^{k}}
\\
  \mel*{\hat{M}_\pm^{(k'+1)}}{\Phi^+}{\hat{M}_\pm^{k+1}} &\simeq \mel*{\hat{M}_\mp^{(k')}}{\Phi^+}{\hat{M}_\mp^{k}}
\end{split}
\end{align}
For the realistic situations studied in the present work, the relative error of this approximation is below $10^{-4}$.
Under the approxmation \eqref{eq:S29}, the matrix elements can be generalised as 
\begin{align}
  \label{eq:S24}
  \mel*{\hat{M}_+^{(k')}}{\Phi^+}{\hat{M}^{(k)}_+} &=
\begin{cases}
0
&
k'\; \mathrm{even}
\\
\Omega_+(M;\zeta)\delta_{k',k+1} + \Omega_-(M;\zeta)\delta_{k',k-1}  
& k' \; \mathrm{odd}
\end{cases}
\\
  \mel*{\hat{M}_-^{(k')}}{\Phi^+}{\hat{M}^{(k)}_-}&=
\begin{cases}
\Omega_+(M;\zeta)\delta_{k',k+1} + \Omega_-(M;\zeta)\delta_{k',k-1}
&
k'\; \mathrm{even}
\\
0
& k' \; \mathrm{odd}
\end{cases}
\end{align}
The interaction Hamiltonian can then be expanded on the derived basis in the form
\begin{align}
  \label{eq:S30} \hat{\mathcal{H}}_I=S_-\sum_{n}
 \mathcal{G}_-\dyad*{\hat{M}_+^{(2n-1)}}{\hat{M}_+^{(2n)}}
+\mathcal{G}_+\dyad*{\hat{M}_+^{(2n+1)}}{\hat{M}_+^{(2n)}} 
+\mathcal{G}_+\dyad*{\hat{M}_-^{(2n)}}{\hat{M}_-^{(2n-1)}}
+\mathcal{G}_-\dyad*{\hat{M}_-^{(2n)}}{\hat{M}_-^{(2n+1)}}
+\mathrm{H.c.}
\end{align}

where the coupling rates are given by $\mathcal{G}_\pm:=\frac{2+\sqrt{2}}{3\pi}\mathcal{A}_\zeta\Omega_\pm$, suppressing explicit dependence of $\Omega_\pm$ on $M$ and $\zeta$.

\section{Quadrupolar inhomogeneities}
\label{sec:quadr-inhom}
In the presence of quadrupolar inhomogeneities, the term $H_\mathrm{Q}^0=\sum_j \Delta_\mathrm{Q}^j(I_z^j)^2$ in Eq.~\eqref{eq:S8} will be non-zero, and the quadrupolar shift, $\Delta_\mathrm{Q}^j$ will be described by a statistical distribution over all the nuclei. This term will be carried onto the interaction Hamiltonian in Eq.~\eqref{eq:S20}, which then becomes $\hat{\mathcal{H}}_I'=\hat{\mathcal{H}}_I+H_Q^0$. To study the effect of this, we consider a fully polarised initial nuclear state, such that the dynamics in the absence of inhomogeneities is spanned by the nuclear states $\ket*{\mathbf{0}}$ and $\ket*{\mathbf{1}}:=\ket*{\hat{\mathbf{0}}_+^{(1)};\zeta}$. The ground state, $\ket{\mathbf{0}}$ is an eigenstate of $H_\mathrm{Q}^0$, but the collective excitation  $\ket{\mathbf{1}}$ is not. The effect of $H_\mathrm{Q}^0$ is then to rotate $\ket*{\mathbf{1}}$ into a set of orthogonal collective excitations, which do not interact with the electron through $\mathcal{H}_I$. To demonstrate this, we define an orthonormal basis for $\mathbb{C}^N$, $\{\nu_\alpha|\alpha=1,\cdots,N\}$, such that $\sum_j \nu_{\alpha,j}^*\nu_{\beta,j}=\delta_{\alpha\beta}$. We choose the first vector to be $\nu_{1,j}=a_j/\sqrt{\sum_j a_j^2}$. These vectors can then be mapped onto a complete basis of spin waves with Zeeman energy $\zeta\omega_\mathrm{Z}^n$, 
\begin{align}
  \label{eq:S11}
  \ket{\alpha}:=\sum_j \nu_{\alpha,j}\ket{\mathbf{0};(\zeta,j)},
\end{align}
such that $\ket{\mathbf{1}}$ corresponds to $\ket{\alpha=1}$. Of all these spin waves, only $\ket{\mathbf{1}}$ is coupled to $\ket{\mathbf{0}}$ via $\mathcal{H}_I$:
\begin{align}
  \label{eq:S21}
  \mel{\mathbf{0}}{\Phi_\zeta^-}{\alpha}=P_-^{(\zeta)}(-I+\zeta)\sum_{j} a_{j} \nu_{\alpha,j}=\delta_{\alpha,1} P_-^{(\zeta)}(-I+\zeta)\sqrt{\sum_j a_j^2}.
\end{align}
and $\mel{\mathbf{0}}{\Phi_\zeta^+}{\alpha}=\mel{\mathbf{\beta}}{\Phi_\zeta^\pm}{\alpha}=0$.
The diffusion rate from $\ket{\mathbf{1}}$ into this dark subspace, $\gamma$, is approximated by calculating the time evolution of $\ket{\mathbf{1}}$ under $H_\mathrm{Q}^0$ and projecting back onto $\ket{\mathbf{1}}$:
\begin{align}
  \label{eq:S25}
  \mel{\mathbf{1}}{e^{-iH_\mathrm{Q}^0 t}}{\mathbf{1}}=\frac{\sum_j a_j^2 e^{-i\Delta_\mathrm{Q}^j \zeta^2 t}}{\sum_j a_j^2}.
\end{align}
Assuming statistical independence of $a_j$ and $\Delta_\mathrm{Q}^j$ and taking the ensemble distribution of $\Delta_\mathrm{Q}^j$ as the normal distribution $p(\Delta_\mathrm{Q})=e^{-\Delta_\mathrm{Q}/(2\sigma^2)}/\sqrt{2\pi\sigma^2}$, we find
\begin{align}
  \label{eq:S31}
  \mel{\mathbf{1}}{e^{-iH_\mathrm{Q}^0 t}}{\mathbf{1}}\simeq \int_{-\infty}^{\infty} \dd{\Delta_\mathrm{Q}} p(\Delta_\mathrm{Q}) e^{-i \Delta_\mathrm{Q}\zeta^2 t} = e^{-\frac{1}{2}(\zeta^2\sigma)^2 t^2},
\end{align}
such that the $1/e$ decay time for the population, $\abs*{\mel{\mathbf{1}}{e^{-iH_\mathrm{Q}^0 t}}{\mathbf{1}}}^2$, is $\gamma=1/(\zeta^2\sigma)$.

To include this effect in the dynamical evolution when assessing the read/write error, we note that the electron-nuclear state $\ket{\phi}\ket{\alpha}$ with $\ket{\alpha}\neq\ket{\mathbf{1}}$ is an eigenstate of the transfer-generating interaction, $\hat{\mathcal{H}}_I$ and thus fully equivalent to $\ket{\phi}\ket{\mathbf{0}}$ when studying the retrieved state of the electron~\cite{SIkurucz2009qubit}.


\begin{thebibliography}{28}%
\makeatletter
\providecommand \@ifxundefined [1]{%
 \@ifx{#1\undefined}
}%
\providecommand \@ifnum [1]{%
 \ifnum #1\expandafter \@firstoftwo
 \else \expandafter \@secondoftwo
 \fi
}%
\providecommand \@ifx [1]{%
 \ifx #1\expandafter \@firstoftwo
 \else \expandafter \@secondoftwo
 \fi
}%
\providecommand \natexlab [1]{#1}%
\providecommand \enquote  [1]{``#1''}%
\providecommand \bibnamefont  [1]{#1}%
\providecommand \bibfnamefont [1]{#1}%
\providecommand \citenamefont [1]{#1}%
\providecommand \href@noop [0]{\@secondoftwo}%
\providecommand \href [0]{\begingroup \@sanitize@url \@href}%
\providecommand \@href[1]{\@@startlink{#1}\@@href}%
\providecommand \@@href[1]{\endgroup#1\@@endlink}%
\providecommand \@sanitize@url [0]{\catcode `\\12\catcode `\$12\catcode
  `\&12\catcode `\#12\catcode `\^12\catcode `\_12\catcode `\%12\relax}%
\providecommand \@@startlink[1]{}%
\providecommand \@@endlink[0]{}%
\providecommand \url  [0]{\begingroup\@sanitize@url \@url }%
\providecommand \@url [1]{\endgroup\@href {#1}{\urlprefix }}%
\providecommand \urlprefix  [0]{URL }%
\providecommand \Eprint [0]{\href }%
\providecommand \doibase [0]{http://dx.doi.org/}%
\providecommand \selectlanguage [0]{\@gobble}%
\providecommand \bibinfo  [0]{\@secondoftwo}%
\providecommand \bibfield  [0]{\@secondoftwo}%
\providecommand \translation [1]{[#1]}%
\providecommand \BibitemOpen [0]{}%
\providecommand \bibitemStop [0]{}%
\providecommand \bibitemNoStop [0]{.\EOS\space}%
\providecommand \EOS [0]{\spacefactor3000\relax}%
\providecommand \BibitemShut  [1]{\csname bibitem#1\endcsname}%
\let\auto@bib@innerbib\@empty
\bibitem [{\citenamefont {Briegel}\ \emph {et~al.}(1998)\citenamefont
  {Briegel}, \citenamefont {D{\"u}r}, \citenamefont {Cirac},\ and\
  \citenamefont {Zoller}}]{briegel1998quantum}%
  \BibitemOpen
  \bibfield  {author} {\bibinfo {author} {\bibfnamefont {H.-J.}\ \bibnamefont
  {Briegel}}, \bibinfo {author} {\bibfnamefont {W.}~\bibnamefont {D{\"u}r}},
  \bibinfo {author} {\bibfnamefont {J.~I.}\ \bibnamefont {Cirac}}, \ and\
  \bibinfo {author} {\bibfnamefont {P.}~\bibnamefont {Zoller}},\ }\href@noop {}
  {\bibfield  {journal} {\bibinfo  {journal} {Physical Review Letters}\
  }\textbf {\bibinfo {volume} {81}},\ \bibinfo {pages} {5932} (\bibinfo {year}
  {1998})}\BibitemShut {NoStop}%
\bibitem [{\citenamefont {Kimble}(2008)}]{kimble2008quantum}%
  \BibitemOpen
  \bibfield  {author} {\bibinfo {author} {\bibfnamefont {H.~J.}\ \bibnamefont
  {Kimble}},\ }\href@noop {} {\bibfield  {journal} {\bibinfo  {journal}
  {Nature}\ }\textbf {\bibinfo {volume} {453}},\ \bibinfo {pages} {1023}
  (\bibinfo {year} {2008})}\BibitemShut {NoStop}%
\bibitem [{\citenamefont {Kozhekin}\ \emph {et~al.}(2000)\citenamefont
  {Kozhekin}, \citenamefont {M{\o}lmer},\ and\ \citenamefont
  {Polzik}}]{kozhekin2000quantum}%
  \BibitemOpen
  \bibfield  {author} {\bibinfo {author} {\bibfnamefont {A.}~\bibnamefont
  {Kozhekin}}, \bibinfo {author} {\bibfnamefont {K.}~\bibnamefont {M{\o}lmer}},
  \ and\ \bibinfo {author} {\bibfnamefont {E.}~\bibnamefont {Polzik}},\
  }\href@noop {} {\bibfield  {journal} {\bibinfo  {journal} {Physical Review
  A}\ }\textbf {\bibinfo {volume} {62}},\ \bibinfo {pages} {033809} (\bibinfo
  {year} {2000})}\BibitemShut {NoStop}%
\bibitem [{\citenamefont {Julsgaard}\ \emph {et~al.}(2004)\citenamefont
  {Julsgaard}, \citenamefont {Sherson}, \citenamefont {Cirac}, \citenamefont
  {Fiur{\'a}{\v{s}}ek},\ and\ \citenamefont
  {Polzik}}]{julsgaard2004experimental}%
  \BibitemOpen
  \bibfield  {author} {\bibinfo {author} {\bibfnamefont {B.}~\bibnamefont
  {Julsgaard}}, \bibinfo {author} {\bibfnamefont {J.}~\bibnamefont {Sherson}},
  \bibinfo {author} {\bibfnamefont {J.~I.}\ \bibnamefont {Cirac}}, \bibinfo
  {author} {\bibfnamefont {J.}~\bibnamefont {Fiur{\'a}{\v{s}}ek}}, \ and\
  \bibinfo {author} {\bibfnamefont {E.~S.}\ \bibnamefont {Polzik}},\
  }\href@noop {} {\bibfield  {journal} {\bibinfo  {journal} {Nature}\ }\textbf
  {\bibinfo {volume} {432}},\ \bibinfo {pages} {482} (\bibinfo {year}
  {2004})}\BibitemShut {NoStop}%
\bibitem [{\citenamefont {Choi}\ \emph {et~al.}(2010)\citenamefont {Choi},
  \citenamefont {Goban}, \citenamefont {Papp}, \citenamefont {Van~Enk},\ and\
  \citenamefont {Kimble}}]{choi2010entanglement}%
  \BibitemOpen
  \bibfield  {author} {\bibinfo {author} {\bibfnamefont {K.}~\bibnamefont
  {Choi}}, \bibinfo {author} {\bibfnamefont {A.}~\bibnamefont {Goban}},
  \bibinfo {author} {\bibfnamefont {S.}~\bibnamefont {Papp}}, \bibinfo {author}
  {\bibfnamefont {S.}~\bibnamefont {Van~Enk}}, \ and\ \bibinfo {author}
  {\bibfnamefont {H.}~\bibnamefont {Kimble}},\ }\href@noop {} {\bibfield
  {journal} {\bibinfo  {journal} {Nature}\ }\textbf {\bibinfo {volume} {468}},\
  \bibinfo {pages} {412} (\bibinfo {year} {2010})}\BibitemShut {NoStop}%
\bibitem [{\citenamefont {Hedges}\ \emph {et~al.}(2010)\citenamefont {Hedges},
  \citenamefont {Longdell}, \citenamefont {Li},\ and\ \citenamefont
  {Sellars}}]{hedges2010efficient}%
  \BibitemOpen
  \bibfield  {author} {\bibinfo {author} {\bibfnamefont {M.~P.}\ \bibnamefont
  {Hedges}}, \bibinfo {author} {\bibfnamefont {J.~J.}\ \bibnamefont
  {Longdell}}, \bibinfo {author} {\bibfnamefont {Y.}~\bibnamefont {Li}}, \ and\
  \bibinfo {author} {\bibfnamefont {M.~J.}\ \bibnamefont {Sellars}},\
  }\href@noop {} {\bibfield  {journal} {\bibinfo  {journal} {Nature}\ }\textbf
  {\bibinfo {volume} {465}},\ \bibinfo {pages} {1052} (\bibinfo {year}
  {2010})}\BibitemShut {NoStop}%
\bibitem [{\citenamefont {Zhao}\ \emph {et~al.}(2009)\citenamefont {Zhao},
  \citenamefont {Dudin}, \citenamefont {Jenkins}, \citenamefont {Campbell},
  \citenamefont {Matsukevich}, \citenamefont {Kennedy},\ and\ \citenamefont
  {Kuzmich}}]{zhao2009long}%
  \BibitemOpen
  \bibfield  {author} {\bibinfo {author} {\bibfnamefont {R.}~\bibnamefont
  {Zhao}}, \bibinfo {author} {\bibfnamefont {Y.}~\bibnamefont {Dudin}},
  \bibinfo {author} {\bibfnamefont {S.}~\bibnamefont {Jenkins}}, \bibinfo
  {author} {\bibfnamefont {C.}~\bibnamefont {Campbell}}, \bibinfo {author}
  {\bibfnamefont {D.}~\bibnamefont {Matsukevich}}, \bibinfo {author}
  {\bibfnamefont {T.}~\bibnamefont {Kennedy}}, \ and\ \bibinfo {author}
  {\bibfnamefont {A.}~\bibnamefont {Kuzmich}},\ }\href@noop {} {\bibfield
  {journal} {\bibinfo  {journal} {Nature Physics}\ }\textbf {\bibinfo {volume}
  {5}},\ \bibinfo {pages} {100} (\bibinfo {year} {2009})}\BibitemShut {NoStop}%
\bibitem [{\citenamefont {Kielpinski}\ \emph {et~al.}(2001)\citenamefont
  {Kielpinski}, \citenamefont {Meyer}, \citenamefont {Rowe}, \citenamefont
  {Sackett}, \citenamefont {Itano}, \citenamefont {Monroe},\ and\ \citenamefont
  {Wineland}}]{kielpinski2001decoherence}%
  \BibitemOpen
  \bibfield  {author} {\bibinfo {author} {\bibfnamefont {D.}~\bibnamefont
  {Kielpinski}}, \bibinfo {author} {\bibfnamefont {V.}~\bibnamefont {Meyer}},
  \bibinfo {author} {\bibfnamefont {M.}~\bibnamefont {Rowe}}, \bibinfo {author}
  {\bibfnamefont {C.}~\bibnamefont {Sackett}}, \bibinfo {author} {\bibfnamefont
  {W.~M.}\ \bibnamefont {Itano}}, \bibinfo {author} {\bibfnamefont
  {C.}~\bibnamefont {Monroe}}, \ and\ \bibinfo {author} {\bibfnamefont {D.~J.}\
  \bibnamefont {Wineland}},\ }\href@noop {} {\bibfield  {journal} {\bibinfo
  {journal} {Science}\ }\textbf {\bibinfo {volume} {291}},\ \bibinfo {pages}
  {1013} (\bibinfo {year} {2001})}\BibitemShut {NoStop}%
\bibitem [{\citenamefont {Langer}\ \emph {et~al.}(2005)\citenamefont {Langer},
  \citenamefont {Ozeri}, \citenamefont {Jost}, \citenamefont {Chiaverini},
  \citenamefont {DeMarco}, \citenamefont {Ben-Kish}, \citenamefont {Blakestad},
  \citenamefont {Britton}, \citenamefont {Hume}, \citenamefont {Itano} \emph
  {et~al.}}]{langer2005long}%
  \BibitemOpen
  \bibfield  {author} {\bibinfo {author} {\bibfnamefont {C.}~\bibnamefont
  {Langer}}, \bibinfo {author} {\bibfnamefont {R.}~\bibnamefont {Ozeri}},
  \bibinfo {author} {\bibfnamefont {J.~D.}\ \bibnamefont {Jost}}, \bibinfo
  {author} {\bibfnamefont {J.}~\bibnamefont {Chiaverini}}, \bibinfo {author}
  {\bibfnamefont {B.}~\bibnamefont {DeMarco}}, \bibinfo {author} {\bibfnamefont
  {A.}~\bibnamefont {Ben-Kish}}, \bibinfo {author} {\bibfnamefont
  {R.}~\bibnamefont {Blakestad}}, \bibinfo {author} {\bibfnamefont
  {J.}~\bibnamefont {Britton}}, \bibinfo {author} {\bibfnamefont
  {D.}~\bibnamefont {Hume}}, \bibinfo {author} {\bibfnamefont {W.~M.}\
  \bibnamefont {Itano}},  \emph {et~al.},\ }\href@noop {} {\bibfield  {journal}
  {\bibinfo  {journal} {Physical Review Letters}\ }\textbf {\bibinfo {volume}
  {95}},\ \bibinfo {pages} {060502} (\bibinfo {year} {2005})}\BibitemShut
  {NoStop}%
\bibitem [{\citenamefont {Dutt}\ \emph {et~al.}(2007)\citenamefont {Dutt},
  \citenamefont {Childress}, \citenamefont {Jiang}, \citenamefont {Togan},
  \citenamefont {Maze}, \citenamefont {Jelezko}, \citenamefont {Zibrov},
  \citenamefont {Hemmer},\ and\ \citenamefont {Lukin}}]{dutt2007quantum}%
  \BibitemOpen
  \bibfield  {author} {\bibinfo {author} {\bibfnamefont {M.~G.}\ \bibnamefont
  {Dutt}}, \bibinfo {author} {\bibfnamefont {L.}~\bibnamefont {Childress}},
  \bibinfo {author} {\bibfnamefont {L.}~\bibnamefont {Jiang}}, \bibinfo
  {author} {\bibfnamefont {E.}~\bibnamefont {Togan}}, \bibinfo {author}
  {\bibfnamefont {J.}~\bibnamefont {Maze}}, \bibinfo {author} {\bibfnamefont
  {F.}~\bibnamefont {Jelezko}}, \bibinfo {author} {\bibfnamefont
  {A.}~\bibnamefont {Zibrov}}, \bibinfo {author} {\bibfnamefont
  {P.}~\bibnamefont {Hemmer}}, \ and\ \bibinfo {author} {\bibfnamefont
  {M.}~\bibnamefont {Lukin}},\ }\href@noop {} {\bibfield  {journal} {\bibinfo
  {journal} {Science}\ }\textbf {\bibinfo {volume} {316}},\ \bibinfo {pages}
  {1312} (\bibinfo {year} {2007})}\BibitemShut {NoStop}%
\bibitem [{\citenamefont {Fuchs}\ \emph {et~al.}(2011)\citenamefont {Fuchs},
  \citenamefont {Burkard}, \citenamefont {Klimov},\ and\ \citenamefont
  {Awschalom}}]{fuchs2011quantum}%
  \BibitemOpen
  \bibfield  {author} {\bibinfo {author} {\bibfnamefont {G.}~\bibnamefont
  {Fuchs}}, \bibinfo {author} {\bibfnamefont {G.}~\bibnamefont {Burkard}},
  \bibinfo {author} {\bibfnamefont {P.}~\bibnamefont {Klimov}}, \ and\ \bibinfo
  {author} {\bibfnamefont {D.}~\bibnamefont {Awschalom}},\ }\href@noop {}
  {\bibfield  {journal} {\bibinfo  {journal} {Nature Physics}\ }\textbf
  {\bibinfo {volume} {7}},\ \bibinfo {pages} {789} (\bibinfo {year}
  {2011})}\BibitemShut {NoStop}%
\bibitem [{\citenamefont {Taminiau}\ \emph {et~al.}(2014)\citenamefont
  {Taminiau}, \citenamefont {Cramer}, \citenamefont {van~der Sar},
  \citenamefont {Dobrovitski},\ and\ \citenamefont
  {Hanson}}]{taminiau2014universal}%
  \BibitemOpen
  \bibfield  {author} {\bibinfo {author} {\bibfnamefont {T.~H.}\ \bibnamefont
  {Taminiau}}, \bibinfo {author} {\bibfnamefont {J.}~\bibnamefont {Cramer}},
  \bibinfo {author} {\bibfnamefont {T.}~\bibnamefont {van~der Sar}}, \bibinfo
  {author} {\bibfnamefont {V.~V.}\ \bibnamefont {Dobrovitski}}, \ and\ \bibinfo
  {author} {\bibfnamefont {R.}~\bibnamefont {Hanson}},\ }\href@noop {}
  {\bibfield  {journal} {\bibinfo  {journal} {Nature Nanotechnology}\ }\textbf
  {\bibinfo {volume} {9}},\ \bibinfo {pages} {171} (\bibinfo {year}
  {2014})}\BibitemShut {NoStop}%
\bibitem [{\citenamefont {Taylor}\ \emph
  {et~al.}(2003{\natexlab{a}})\citenamefont {Taylor}, \citenamefont {Marcus},\
  and\ \citenamefont {Lukin}}]{taylor2003long}%
  \BibitemOpen
  \bibfield  {author} {\bibinfo {author} {\bibfnamefont {J.}~\bibnamefont
  {Taylor}}, \bibinfo {author} {\bibfnamefont {C.}~\bibnamefont {Marcus}}, \
  and\ \bibinfo {author} {\bibfnamefont {M.}~\bibnamefont {Lukin}},\
  }\href@noop {} {\bibfield  {journal} {\bibinfo  {journal} {Physical Review
  Letters}\ }\textbf {\bibinfo {volume} {90}},\ \bibinfo {pages} {206803}
  (\bibinfo {year} {2003}{\natexlab{a}})}\BibitemShut {NoStop}%
\bibitem [{\citenamefont {Taylor}\ \emph
  {et~al.}(2003{\natexlab{b}})\citenamefont {Taylor}, \citenamefont
  {Imamoglu},\ and\ \citenamefont {Lukin}}]{taylor2003controlling}%
  \BibitemOpen
  \bibfield  {author} {\bibinfo {author} {\bibfnamefont {J.}~\bibnamefont
  {Taylor}}, \bibinfo {author} {\bibfnamefont {A.}~\bibnamefont {Imamoglu}}, \
  and\ \bibinfo {author} {\bibfnamefont {M.}~\bibnamefont {Lukin}},\
  }\href@noop {} {\bibfield  {journal} {\bibinfo  {journal} {Physical Review
  Letters}\ }\textbf {\bibinfo {volume} {91}},\ \bibinfo {pages} {246802}
  (\bibinfo {year} {2003}{\natexlab{b}})}\BibitemShut {NoStop}%
\bibitem [{\citenamefont {Kurucz}\ \emph {et~al.}(2009)\citenamefont {Kurucz},
  \citenamefont {S{\o}rensen}, \citenamefont {Taylor}, \citenamefont {Lukin},\
  and\ \citenamefont {Fleischhauer}}]{kurucz2009qubit}%
  \BibitemOpen
  \bibfield  {author} {\bibinfo {author} {\bibfnamefont {Z.}~\bibnamefont
  {Kurucz}}, \bibinfo {author} {\bibfnamefont {M.~W.}\ \bibnamefont
  {S{\o}rensen}}, \bibinfo {author} {\bibfnamefont {J.~M.}\ \bibnamefont
  {Taylor}}, \bibinfo {author} {\bibfnamefont {M.~D.}\ \bibnamefont {Lukin}}, \
  and\ \bibinfo {author} {\bibfnamefont {M.}~\bibnamefont {Fleischhauer}},\
  }\href@noop {} {\bibfield  {journal} {\bibinfo  {journal} {Physical Review
  Letters}\ }\textbf {\bibinfo {volume} {103}},\ \bibinfo {pages} {010502}
  (\bibinfo {year} {2009})}\BibitemShut {NoStop}%
\bibitem [{\citenamefont {Chekhovich}\ \emph {et~al.}(2015)\citenamefont
  {Chekhovich}, \citenamefont {Hopkinson}, \citenamefont {Skolnick},\ and\
  \citenamefont {Tartakovskii}}]{chekhovich2015suppression}%
  \BibitemOpen
  \bibfield  {author} {\bibinfo {author} {\bibfnamefont {E.}~\bibnamefont
  {Chekhovich}}, \bibinfo {author} {\bibfnamefont {M.}~\bibnamefont
  {Hopkinson}}, \bibinfo {author} {\bibfnamefont {M.}~\bibnamefont {Skolnick}},
  \ and\ \bibinfo {author} {\bibfnamefont {A.}~\bibnamefont {Tartakovskii}},\
  }\href@noop {} {\bibfield  {journal} {\bibinfo  {journal} {Nature
  communications}\ }\textbf {\bibinfo {volume} {6}},\ \bibinfo {pages} {6348}
  (\bibinfo {year} {2015})}\BibitemShut {NoStop}%
\bibitem [{\citenamefont {W{\"u}st}\ \emph {et~al.}(2016)\citenamefont
  {W{\"u}st}, \citenamefont {Munsch}, \citenamefont {Maier}, \citenamefont
  {Kuhlmann}, \citenamefont {Ludwig}, \citenamefont {Wieck}, \citenamefont
  {Loss}, \citenamefont {Poggio},\ and\ \citenamefont
  {Warburton}}]{wust2016role}%
  \BibitemOpen
  \bibfield  {author} {\bibinfo {author} {\bibfnamefont {G.}~\bibnamefont
  {W{\"u}st}}, \bibinfo {author} {\bibfnamefont {M.}~\bibnamefont {Munsch}},
  \bibinfo {author} {\bibfnamefont {F.}~\bibnamefont {Maier}}, \bibinfo
  {author} {\bibfnamefont {A.~V.}\ \bibnamefont {Kuhlmann}}, \bibinfo {author}
  {\bibfnamefont {A.}~\bibnamefont {Ludwig}}, \bibinfo {author} {\bibfnamefont
  {A.~D.}\ \bibnamefont {Wieck}}, \bibinfo {author} {\bibfnamefont
  {D.}~\bibnamefont {Loss}}, \bibinfo {author} {\bibfnamefont {M.}~\bibnamefont
  {Poggio}}, \ and\ \bibinfo {author} {\bibfnamefont {R.~J.}\ \bibnamefont
  {Warburton}},\ }\href@noop {} {\bibfield  {journal} {\bibinfo  {journal}
  {Nature Nanotechnology}\ }\textbf {\bibinfo {volume} {11}},\ \bibinfo {pages}
  {885} (\bibinfo {year} {2016})}\BibitemShut {NoStop}%
\bibitem [{\citenamefont {Gangloff}\ \emph {et~al.}(2019)\citenamefont
  {Gangloff}, \citenamefont {Ethier-Majcher}, \citenamefont {Lang},
  \citenamefont {Denning}, \citenamefont {Bodey}, \citenamefont {Jackson},
  \citenamefont {Clarke}, \citenamefont {Hugues}, \citenamefont {Le~Gall},\
  and\ \citenamefont {Atat\"ure}}]{gangloff2019magnon}%
  \BibitemOpen
  \bibfield  {author} {\bibinfo {author} {\bibfnamefont {D.~A.}\ \bibnamefont
  {Gangloff}}, \bibinfo {author} {\bibfnamefont {G.}~\bibnamefont
  {Ethier-Majcher}}, \bibinfo {author} {\bibfnamefont {C.}~\bibnamefont
  {Lang}}, \bibinfo {author} {\bibfnamefont {E.~V.}\ \bibnamefont {Denning}},
  \bibinfo {author} {\bibfnamefont {J.~H.}\ \bibnamefont {Bodey}}, \bibinfo
  {author} {\bibfnamefont {D.}~\bibnamefont {Jackson}}, \bibinfo {author}
  {\bibfnamefont {E.}~\bibnamefont {Clarke}}, \bibinfo {author} {\bibfnamefont
  {M.}~\bibnamefont {Hugues}}, \bibinfo {author} {\bibfnamefont
  {C.}~\bibnamefont {Le~Gall}}, \ and\ \bibinfo {author} {\bibfnamefont
  {M.}~\bibnamefont {Atat\"ure}},\ }\href@noop {} {\bibfield  {journal}
  {\bibinfo  {journal} {Science}\ }\textbf {\bibinfo {volume} {364}},\ \bibinfo
  {pages} {62} (\bibinfo {year} {2019})}\BibitemShut {NoStop}%
\bibitem [{\citenamefont {H{\"o}gele}\ \emph {et~al.}(2012)\citenamefont
  {H{\"o}gele}, \citenamefont {Kroner}, \citenamefont {Latta}, \citenamefont
  {Claassen}, \citenamefont {Carusotto}, \citenamefont {Bulutay},\ and\
  \citenamefont {Imamoglu}}]{hogele2012dynamic}%
  \BibitemOpen
  \bibfield  {author} {\bibinfo {author} {\bibfnamefont {A.}~\bibnamefont
  {H{\"o}gele}}, \bibinfo {author} {\bibfnamefont {M.}~\bibnamefont {Kroner}},
  \bibinfo {author} {\bibfnamefont {C.}~\bibnamefont {Latta}}, \bibinfo
  {author} {\bibfnamefont {M.}~\bibnamefont {Claassen}}, \bibinfo {author}
  {\bibfnamefont {I.}~\bibnamefont {Carusotto}}, \bibinfo {author}
  {\bibfnamefont {C.}~\bibnamefont {Bulutay}}, \ and\ \bibinfo {author}
  {\bibfnamefont {A.}~\bibnamefont {Imamoglu}},\ }\href@noop {} {\bibfield
  {journal} {\bibinfo  {journal} {Physical Review Letters}\ }\textbf {\bibinfo
  {volume} {108}},\ \bibinfo {pages} {197403} (\bibinfo {year}
  {2012})}\BibitemShut {NoStop}%
\bibitem [{\citenamefont {Schwartz}\ \emph {et~al.}(2018)\citenamefont
  {Schwartz}, \citenamefont {Scheuer}, \citenamefont {Tratzmiller},
  \citenamefont {M{\"u}ller}, \citenamefont {Chen}, \citenamefont {Dhand},
  \citenamefont {Wang}, \citenamefont {M{\"u}ller}, \citenamefont {Naydenov},
  \citenamefont {Jelezko} \emph {et~al.}}]{schwartz2018robust}%
  \BibitemOpen
  \bibfield  {author} {\bibinfo {author} {\bibfnamefont {I.}~\bibnamefont
  {Schwartz}}, \bibinfo {author} {\bibfnamefont {J.}~\bibnamefont {Scheuer}},
  \bibinfo {author} {\bibfnamefont {B.}~\bibnamefont {Tratzmiller}}, \bibinfo
  {author} {\bibfnamefont {S.}~\bibnamefont {M{\"u}ller}}, \bibinfo {author}
  {\bibfnamefont {Q.}~\bibnamefont {Chen}}, \bibinfo {author} {\bibfnamefont
  {I.}~\bibnamefont {Dhand}}, \bibinfo {author} {\bibfnamefont {Z.-Y.}\
  \bibnamefont {Wang}}, \bibinfo {author} {\bibfnamefont {C.}~\bibnamefont
  {M{\"u}ller}}, \bibinfo {author} {\bibfnamefont {B.}~\bibnamefont
  {Naydenov}}, \bibinfo {author} {\bibfnamefont {F.}~\bibnamefont {Jelezko}},
  \emph {et~al.},\ }\href@noop {} {\bibfield  {journal} {\bibinfo  {journal}
  {Science Advances}\ }\textbf {\bibinfo {volume} {4}},\ \bibinfo {pages}
  {eaat8978} (\bibinfo {year} {2018})}\BibitemShut {NoStop}%
\bibitem [{\citenamefont {Bodey}\ \emph {et~al.}(2019)\citenamefont {Bodey}
  \emph {et~al.}}]{bodey2019ESR}%
  \BibitemOpen
  \bibfield  {author} {\bibinfo {author} {\bibfnamefont {J.~H.}\ \bibnamefont
  {Bodey}} \emph {et~al.},\ }\href@noop {} {\bibfield  {journal} {\bibinfo
  {journal} {in preparation}\ } (\bibinfo {year} {2019})}\BibitemShut {NoStop}%
\bibitem [{\citenamefont {Yao}\ \emph {et~al.}(2006)\citenamefont {Yao},
  \citenamefont {Liu},\ and\ \citenamefont {Sham}}]{yao2006theory}%
  \BibitemOpen
  \bibfield  {author} {\bibinfo {author} {\bibfnamefont {W.}~\bibnamefont
  {Yao}}, \bibinfo {author} {\bibfnamefont {R.-B.}\ \bibnamefont {Liu}}, \ and\
  \bibinfo {author} {\bibfnamefont {L.}~\bibnamefont {Sham}},\ }\href@noop {}
  {\bibfield  {journal} {\bibinfo  {journal} {Physical Review B}\ }\textbf
  {\bibinfo {volume} {74}},\ \bibinfo {pages} {195301} (\bibinfo {year}
  {2006})}\BibitemShut {NoStop}%
\bibitem [{sup()}]{suppmat}%
  \BibitemOpen
  \href@noop {} {\emph {\bibinfo {title} {See Supplemental
  Material}}}\BibitemShut {NoStop}%
\bibitem [{\citenamefont {Bulutay}(2012)}]{bulutay2012quadrupolar}%
  \BibitemOpen
  \bibfield  {author} {\bibinfo {author} {\bibfnamefont {C.}~\bibnamefont
  {Bulutay}},\ }\href@noop {} {\bibfield  {journal} {\bibinfo  {journal}
  {Physical Review B}\ }\textbf {\bibinfo {volume} {85}},\ \bibinfo {pages}
  {115313} (\bibinfo {year} {2012})}\BibitemShut {NoStop}%
\bibitem [{\citenamefont {Yuan}\ \emph {et~al.}(2018)\citenamefont {Yuan},
  \citenamefont {Weyhausen-Brinkmann}, \citenamefont {Mart{\'\i}n-S{\'a}nchez},
  \citenamefont {Piredda}, \citenamefont {K{\v{r}}{\'a}pek}, \citenamefont
  {Huo}, \citenamefont {Huang}, \citenamefont {Schimpf}, \citenamefont
  {Schmidt}, \citenamefont {Edlinger} \emph {et~al.}}]{yuan2018uniaxial}%
  \BibitemOpen
  \bibfield  {author} {\bibinfo {author} {\bibfnamefont {X.}~\bibnamefont
  {Yuan}}, \bibinfo {author} {\bibfnamefont {F.}~\bibnamefont
  {Weyhausen-Brinkmann}}, \bibinfo {author} {\bibfnamefont {J.}~\bibnamefont
  {Mart{\'\i}n-S{\'a}nchez}}, \bibinfo {author} {\bibfnamefont
  {G.}~\bibnamefont {Piredda}}, \bibinfo {author} {\bibfnamefont
  {V.}~\bibnamefont {K{\v{r}}{\'a}pek}}, \bibinfo {author} {\bibfnamefont
  {Y.}~\bibnamefont {Huo}}, \bibinfo {author} {\bibfnamefont {H.}~\bibnamefont
  {Huang}}, \bibinfo {author} {\bibfnamefont {C.}~\bibnamefont {Schimpf}},
  \bibinfo {author} {\bibfnamefont {O.~G.}\ \bibnamefont {Schmidt}}, \bibinfo
  {author} {\bibfnamefont {J.}~\bibnamefont {Edlinger}},  \emph {et~al.},\
  }\href@noop {} {\bibfield  {journal} {\bibinfo  {journal} {Nature
  communications}\ }\textbf {\bibinfo {volume} {9}},\ \bibinfo {pages} {3058}
  (\bibinfo {year} {2018})}\BibitemShut {NoStop}%
\bibitem [{\citenamefont {Flisinski}\ \emph {et~al.}(2010)\citenamefont
  {Flisinski}, \citenamefont {Gerlovin}, \citenamefont {Ignatiev},
  \citenamefont {Petrov}, \citenamefont {Verbin}, \citenamefont {Yakovlev},
  \citenamefont {Reuter}, \citenamefont {Wieck},\ and\ \citenamefont
  {Bayer}}]{flisinski2010optically}%
  \BibitemOpen
  \bibfield  {author} {\bibinfo {author} {\bibfnamefont {K.}~\bibnamefont
  {Flisinski}}, \bibinfo {author} {\bibfnamefont {I.~Y.}\ \bibnamefont
  {Gerlovin}}, \bibinfo {author} {\bibfnamefont {I.}~\bibnamefont {Ignatiev}},
  \bibinfo {author} {\bibfnamefont {M.~Y.}\ \bibnamefont {Petrov}}, \bibinfo
  {author} {\bibfnamefont {S.~Y.}\ \bibnamefont {Verbin}}, \bibinfo {author}
  {\bibfnamefont {D.}~\bibnamefont {Yakovlev}}, \bibinfo {author}
  {\bibfnamefont {D.}~\bibnamefont {Reuter}}, \bibinfo {author} {\bibfnamefont
  {A.}~\bibnamefont {Wieck}}, \ and\ \bibinfo {author} {\bibfnamefont
  {M.}~\bibnamefont {Bayer}},\ }\href@noop {} {\bibfield  {journal} {\bibinfo
  {journal} {Physical Review B}\ }\textbf {\bibinfo {volume} {82}},\ \bibinfo
  {pages} {081308} (\bibinfo {year} {2010})}\BibitemShut {NoStop}%
\bibitem [{\citenamefont {Chekhovich}\ \emph {et~al.}(2017)\citenamefont
  {Chekhovich}, \citenamefont {Ulhaq}, \citenamefont {Zallo}, \citenamefont
  {Ding}, \citenamefont {Schmidt},\ and\ \citenamefont
  {Skolnick}}]{chekhovich2017measurement}%
  \BibitemOpen
  \bibfield  {author} {\bibinfo {author} {\bibfnamefont {E.}~\bibnamefont
  {Chekhovich}}, \bibinfo {author} {\bibfnamefont {A.}~\bibnamefont {Ulhaq}},
  \bibinfo {author} {\bibfnamefont {E.}~\bibnamefont {Zallo}}, \bibinfo
  {author} {\bibfnamefont {F.}~\bibnamefont {Ding}}, \bibinfo {author}
  {\bibfnamefont {O.}~\bibnamefont {Schmidt}}, \ and\ \bibinfo {author}
  {\bibfnamefont {M.}~\bibnamefont {Skolnick}},\ }\href@noop {} {\bibfield
  {journal} {\bibinfo  {journal} {Nature materials}\ }\textbf {\bibinfo
  {volume} {16}},\ \bibinfo {pages} {982} (\bibinfo {year} {2017})}\BibitemShut
  {NoStop}%
\bibitem [{\citenamefont {Chekhovich}(2017)}]{chekhovich2017decoherence}%
  \BibitemOpen
  \bibfield  {author} {\bibinfo {author} {\bibfnamefont {E.~A.}\ \bibnamefont
  {Chekhovich}},\ }in\ \href@noop {} {\emph {\bibinfo {booktitle} {Journal of
  Physics: Conference Series}}},\ Vol.\ \bibinfo {volume} {864}\ (\bibinfo
  {organization} {IOP Publishing},\ \bibinfo {year} {2017})\ p.\ \bibinfo
  {pages} {012080}\BibitemShut {NoStop}%
\end{thebibliography}

\begin{thebibliography}{6}%
\makeatletter
\providecommand \@ifxundefined [1]{%
 \@ifx{#1\undefined}
}%
\providecommand \@ifnum [1]{%
 \ifnum #1\expandafter \@firstoftwo
 \else \expandafter \@secondoftwo
 \fi
}%
\providecommand \@ifx [1]{%
 \ifx #1\expandafter \@firstoftwo
 \else \expandafter \@secondoftwo
 \fi
}%
\providecommand \natexlab [1]{#1}%
\providecommand \enquote  [1]{``#1''}%
\providecommand \bibnamefont  [1]{#1}%
\providecommand \bibfnamefont [1]{#1}%
\providecommand \citenamefont [1]{#1}%
\providecommand \href@noop [0]{\@secondoftwo}%
\providecommand \href [0]{\begingroup \@sanitize@url \@href}%
\providecommand \@href[1]{\@@startlink{#1}\@@href}%
\providecommand \@@href[1]{\endgroup#1\@@endlink}%
\providecommand \@sanitize@url [0]{\catcode `\\12\catcode `\$12\catcode
  `\&12\catcode `\#12\catcode `\^12\catcode `\_12\catcode `\%12\relax}%
\providecommand \@@startlink[1]{}%
\providecommand \@@endlink[0]{}%
\providecommand \url  [0]{\begingroup\@sanitize@url \@url }%
\providecommand \@url [1]{\endgroup\@href {#1}{\urlprefix }}%
\providecommand \urlprefix  [0]{URL }%
\providecommand \Eprint [0]{\href }%
\providecommand \doibase [0]{http://dx.doi.org/}%
\providecommand \selectlanguage [0]{\@gobble}%
\providecommand \bibinfo  [0]{\@secondoftwo}%
\providecommand \bibfield  [0]{\@secondoftwo}%
\providecommand \translation [1]{[#1]}%
\providecommand \BibitemOpen [0]{}%
\providecommand \bibitemStop [0]{}%
\providecommand \bibitemNoStop [0]{.\EOS\space}%
\providecommand \EOS [0]{\spacefactor3000\relax}%
\providecommand \BibitemShut  [1]{\csname bibitem#1\endcsname}%
\let\auto@bib@innerbib\@empty
\bibitem [{\citenamefont {Bulutay}(2012)}]{SIbulutay2012quadrupolar}%
  \BibitemOpen
  \bibfield  {author} {\bibinfo {author} {\bibfnamefont {C.}~\bibnamefont
  {Bulutay}},\ }\href@noop {} {\bibfield  {journal} {\bibinfo  {journal}
  {Physical Review B}\ }\textbf {\bibinfo {volume} {85}},\ \bibinfo {pages}
  {115313} (\bibinfo {year} {2012})}\BibitemShut {NoStop}%
\bibitem [{\citenamefont {Urbaszek}\ \emph {et~al.}(2013)\citenamefont
  {Urbaszek}, \citenamefont {Marie}, \citenamefont {Amand}, \citenamefont
  {Krebs}, \citenamefont {Voisin}, \citenamefont {Maletinsky}, \citenamefont
  {H{\"o}gele},\ and\ \citenamefont {Imamoglu}}]{SIurbaszek2013nuclear}%
  \BibitemOpen
  \bibfield  {author} {\bibinfo {author} {\bibfnamefont {B.}~\bibnamefont
  {Urbaszek}}, \bibinfo {author} {\bibfnamefont {X.}~\bibnamefont {Marie}},
  \bibinfo {author} {\bibfnamefont {T.}~\bibnamefont {Amand}}, \bibinfo
  {author} {\bibfnamefont {O.}~\bibnamefont {Krebs}}, \bibinfo {author}
  {\bibfnamefont {P.}~\bibnamefont {Voisin}}, \bibinfo {author} {\bibfnamefont
  {P.}~\bibnamefont {Maletinsky}}, \bibinfo {author} {\bibfnamefont
  {A.}~\bibnamefont {H{\"o}gele}}, \ and\ \bibinfo {author} {\bibfnamefont
  {A.}~\bibnamefont {Imamoglu}},\ }\href@noop {} {\bibfield  {journal}
  {\bibinfo  {journal} {Reviews of Modern Physics}\ }\textbf {\bibinfo {volume}
  {85}},\ \bibinfo {pages} {79} (\bibinfo {year} {2013})}\BibitemShut {NoStop}%
\bibitem [{\citenamefont {Gangloff}\ \emph {et~al.}(2019)\citenamefont
  {Gangloff}, \citenamefont {Ethier-Majcher}, \citenamefont {Lang},
  \citenamefont {Denning}, \citenamefont {Bodey}, \citenamefont {Jackson},
  \citenamefont {Clarke}, \citenamefont {Hugues}, \citenamefont {Le~Gall},\
  and\ \citenamefont {Atat\"ure}}]{SIgangloff2019magnon}%
  \BibitemOpen
  \bibfield  {author} {\bibinfo {author} {\bibfnamefont {D.~A.}\ \bibnamefont
  {Gangloff}}, \bibinfo {author} {\bibfnamefont {G.}~\bibnamefont
  {Ethier-Majcher}}, \bibinfo {author} {\bibfnamefont {C.}~\bibnamefont
  {Lang}}, \bibinfo {author} {\bibfnamefont {E.~V.}\ \bibnamefont {Denning}},
  \bibinfo {author} {\bibfnamefont {J.~H.}\ \bibnamefont {Bodey}}, \bibinfo
  {author} {\bibfnamefont {D.}~\bibnamefont {Jackson}}, \bibinfo {author}
  {\bibfnamefont {E.}~\bibnamefont {Clarke}}, \bibinfo {author} {\bibfnamefont
  {M.}~\bibnamefont {Hugues}}, \bibinfo {author} {\bibfnamefont
  {C.}~\bibnamefont {Le~Gall}}, \ and\ \bibinfo {author} {\bibfnamefont
  {M.}~\bibnamefont {Atat\"ure}},\ }\href@noop {} {\bibfield  {journal}
  {\bibinfo  {journal} {Science}\ }\textbf {\bibinfo {volume} {364}},\ \bibinfo
  {pages} {62} (\bibinfo {year} {2019})}\BibitemShut {NoStop}%
\bibitem [{\citenamefont {Huo}\ \emph {et~al.}(2014)\citenamefont {Huo},
  \citenamefont {Witek}, \citenamefont {Kumar}, \citenamefont {Cardenas},
  \citenamefont {Zhang}, \citenamefont {Akopian}, \citenamefont {Singh},
  \citenamefont {Zallo}, \citenamefont {Grifone}, \citenamefont {Kriegner}
  \emph {et~al.}}]{SIhuo2014light}%
  \BibitemOpen
  \bibfield  {author} {\bibinfo {author} {\bibfnamefont {Y.}~\bibnamefont
  {Huo}}, \bibinfo {author} {\bibfnamefont {B.}~\bibnamefont {Witek}}, \bibinfo
  {author} {\bibfnamefont {S.}~\bibnamefont {Kumar}}, \bibinfo {author}
  {\bibfnamefont {J.}~\bibnamefont {Cardenas}}, \bibinfo {author}
  {\bibfnamefont {J.}~\bibnamefont {Zhang}}, \bibinfo {author} {\bibfnamefont
  {N.}~\bibnamefont {Akopian}}, \bibinfo {author} {\bibfnamefont
  {R.}~\bibnamefont {Singh}}, \bibinfo {author} {\bibfnamefont
  {E.}~\bibnamefont {Zallo}}, \bibinfo {author} {\bibfnamefont
  {R.}~\bibnamefont {Grifone}}, \bibinfo {author} {\bibfnamefont
  {D.}~\bibnamefont {Kriegner}},  \emph {et~al.},\ }\href@noop {} {\bibfield
  {journal} {\bibinfo  {journal} {Nature Physics}\ }\textbf {\bibinfo {volume}
  {10}},\ \bibinfo {pages} {46} (\bibinfo {year} {2014})}\BibitemShut {NoStop}%
\bibitem [{\citenamefont {Schwartz}\ \emph {et~al.}(2018)\citenamefont
  {Schwartz}, \citenamefont {Scheuer}, \citenamefont {Tratzmiller},
  \citenamefont {M{\"u}ller}, \citenamefont {Chen}, \citenamefont {Dhand},
  \citenamefont {Wang}, \citenamefont {M{\"u}ller}, \citenamefont {Naydenov},
  \citenamefont {Jelezko} \emph {et~al.}}]{SIschwartz2018robust}%
  \BibitemOpen
  \bibfield  {author} {\bibinfo {author} {\bibfnamefont {I.}~\bibnamefont
  {Schwartz}}, \bibinfo {author} {\bibfnamefont {J.}~\bibnamefont {Scheuer}},
  \bibinfo {author} {\bibfnamefont {B.}~\bibnamefont {Tratzmiller}}, \bibinfo
  {author} {\bibfnamefont {S.}~\bibnamefont {M{\"u}ller}}, \bibinfo {author}
  {\bibfnamefont {Q.}~\bibnamefont {Chen}}, \bibinfo {author} {\bibfnamefont
  {I.}~\bibnamefont {Dhand}}, \bibinfo {author} {\bibfnamefont {Z.-Y.}\
  \bibnamefont {Wang}}, \bibinfo {author} {\bibfnamefont {C.}~\bibnamefont
  {M{\"u}ller}}, \bibinfo {author} {\bibfnamefont {B.}~\bibnamefont
  {Naydenov}}, \bibinfo {author} {\bibfnamefont {F.}~\bibnamefont {Jelezko}},
  \emph {et~al.},\ }\href@noop {} {\bibfield  {journal} {\bibinfo  {journal}
  {Science advances}\ }\textbf {\bibinfo {volume} {4}},\ \bibinfo {pages}
  {eaat8978} (\bibinfo {year} {2018})}\BibitemShut {NoStop}%
\bibitem [{\citenamefont {Kurucz}\ \emph {et~al.}(2009)\citenamefont {Kurucz},
  \citenamefont {S{\o}rensen}, \citenamefont {Taylor}, \citenamefont {Lukin},\
  and\ \citenamefont {Fleischhauer}}]{SIkurucz2009qubit}%
  \BibitemOpen
  \bibfield  {author} {\bibinfo {author} {\bibfnamefont {Z.}~\bibnamefont
  {Kurucz}}, \bibinfo {author} {\bibfnamefont {M.~W.}\ \bibnamefont
  {S{\o}rensen}}, \bibinfo {author} {\bibfnamefont {J.~M.}\ \bibnamefont
  {Taylor}}, \bibinfo {author} {\bibfnamefont {M.~D.}\ \bibnamefont {Lukin}}, \
  and\ \bibinfo {author} {\bibfnamefont {M.}~\bibnamefont {Fleischhauer}},\
  }\href@noop {} {\bibfield  {journal} {\bibinfo  {journal} {Physical review
  letters}\ }\textbf {\bibinfo {volume} {103}},\ \bibinfo {pages} {010502}
  (\bibinfo {year} {2009})}\BibitemShut {NoStop}%
\end{thebibliography}
\end{document}